\documentclass[11pt]{article}
 \usepackage[margin=1.5cm]{geometry}
 \usepackage[parfill]{parskip} 
\usepackage{amsmath,amssymb}
\usepackage[numbers]{natbib}

\usepackage{graphicx}
\usepackage{dcolumn}
\usepackage{bm}

\usepackage[utf8]{inputenc} 
\usepackage[T1]{fontenc}
\usepackage{color}
\usepackage[normalem]{ulem} 
\usepackage{textcomp} 
\usepackage{blindtext}
\usepackage{tikz,float,subfig} 

\usetikzlibrary{shapes,shadows,arrows,snakes,angles,quotes,plotmarks}
\tikzstyle{line} = [draw, -latex']
\tikzstyle{block} = [draw, rectangle, fill=blue!50, text centered, minimum height=10mm, node distance=10em]

\usepackage{multirow,booktabs} 

\DeclareFontFamily{T1}{calligra}{} \DeclareFontShape{T1}{calligra}{m}{n}{<->s*[1.44]callig15}{} \DeclareMathAlphabet\mathzapf {T1}{pzc} {mb} {it}
 \DeclareMathOperator{\Pe}{Pe} \DeclareMathOperator{\Da}{Da} \DeclareMathOperator{\Var}{Var} \DeclareMathOperator{\eff}{eff}

\newcommand{\chg}[2][black]{{\color{#1}#2}}
\newcommand\rem[1]{}   

\title{Advection-dominated transport past isolated disordered sinks: \\ stepping beyond homogenization}

\author{
George F. Price$^{1}$, Igor L. Chernyavsky$^{1,2}$ and Oliver E. Jensen$^{1}$ \\
$^{1}$Department of Mathematics, University of Manchester, UK\\
$^{2}$ Maternal and Fetal Health Research Centre, University of Manchester, UK}

\begin{document}
\maketitle

\begin{abstract}
We investigate the transport of a solute past isolated sinks in a bounded domain when advection is dominant over diffusion, evaluating the effectiveness of homogenization approximations when sinks are distributed uniformly randomly in space.  Corrections to such approximations can be non-local, non-smooth and non-Gaussian, depending on the physical parameters (a P\'eclet number Pe, assumed large, and a Damk\"ohler number Da) and the compactness of the sinks.  In one spatial dimension, solute distributions develop a staircase structure for large $\Pe$, with corrections being better described with credible intervals than with traditional moments.  In two and three dimensions, solute distributions are near-singular at each sink (and regularized by sink size), but their moments can be smooth as a result of ensemble averaging over variable sink locations.  We approximate corrections to a homogenization approximation using a moment-expansion method, replacing the Green's function by its free-space form, and test predictions against simulation.  We show how, in two or three dimensions, the leading-order impact of disorder can be captured in a homogenization approximation for the ensemble mean concentration through a modification to $\Da$ that grows with diminishing sink size.
\end{abstract}


\section{Introduction}

Transport processes in many natural systems take place in spatially disordered domains.  In many instances, these processes can be adequately described by averaging procedures, Darcy's law describing flow in random porous media being a well-known example \cite{Rubinstein1989Flow}.  However it is important to understand the impact of disorder, particularly in instances where disorder has a significant influence (for example in explaining breakthrough effects, whereby solute is carried rapidly along a small number of high-flow paths through a random porous medium \cite{Berkowitz2006Modeling}).  The present study contributes to this effort by characterising the impact of spatial disorder on the uptake of a solute that is advected past distributions of isolated sinks.  This problem is loosely motivated by transport of maternal blood in the intervillous space of the human placenta \citep{ChernDrydenJensen12} but is posed here in more general terms.

A common assumption that is exploited in order to describe \chg{transport} in media with complex microstructure is to assume periodicity at the microscale \citep{Allaire2007Homogenization, Davit2013Homogenization, hornung1991, mauri1991}.  This allows an asymptotic two-scale expansion to be developed, with a unit-cell problem (with periodic boundary conditions) being solved in order to provide a description of slowly varying (homogenized) variables at the macroscale.  While this approach has been extended to accommodate slow spatial variation of the microscale field \cite{Bruna2015Diffusion, Dalwadi2015Understanding, ray2015} and developed for a variety of applications \cite{Chapman2008Multiscale, Dalwadi2018Upscaling, RamirezTorres2018asymptotic, Dalwadi2020systematic, Mahiout2020Homogenization, Piatnitski2017Homogenization}, it is less adaptable to situations where the microscale exhibits appreciable spatial disorder.  \chg{Approaches currently adopted in such instances include formal methods of stochastic homogenization \cite{gloria2015}, spatial averaging techniques \cite{quintard1993} or simulations using random microstructures realized within periodic unit cells \cite{Printsypar2019influence}. }

A spatially disordered medium can be characterised as a random field with prescribed statistical properties. The `forward' problem that we address here seeks to understand how these properties map to the statistical properties of the concentration field of a solute as it passes through the medium.  This map is mediated by physical processes embodied in a partial differential equation (in the present instance, a linear advection-diffusion-reaction equation).  The primary question addressed by a homogenization approximation is how to translate the first moment of the sink density to the first moment of the associated concentration field (where first moments are ensemble averages).  More refined questions address the impact of spatial disorder, captured in the second moment (covariance) of the sink density, on the mean and covariance of the concentration field.  Provided solute fluctuations are bounded in an appropriate sense, these corrections can be evaluated by perturbation around the leading-order homogenization approximation, as we illustrate below, and as demonstrated previously by \citet{Dagan1984Solute}, \citet{Cushman2002primer}, \citet{ChernDrydenJensen12}, \citet{RussellJensenGalla16} and \citet{Russell2020Homogenization}.  If fluctuations become sufficiently large, or if distributions become strongly non-Gaussian, higher moments (or even full probability distributions) of the solute field may need to be evaluated.  

Homogenization approximations exploit the separation of lengthscales between the microscale and the macroscale.  However, when considering solute uptake at isolated sinks, a further lengthscale needs consideration.  The microscale involves two lengthscales, an intersink distance $\rho$ (assumed small compared to the overall size of the domain) and a sink size $\varsigma$.  As $\varsigma$ becomes vanishingly small with respect to $\rho$, over the shortest lengthscales, diffusion can be expected to dominate advection in the neighbourhood of sinks, and the concentration field can be expected to be described locally by the solution of a diffusion equation in the neighbourhood of a point source.  In one dimension (1D), this leads to a concentration field with a staircase structure, with a thin diffusive boundary layer forming upstream of each sink \cite{Russell2020Homogenization}.  In two and three spatial dimensions (2D and 3D), large solute gradients surround the sink, and the concentration field grows in magnitude proportionally to $\log (\rho/\varsigma)$ and $\rho/\varsigma$ respectively.  This effect amplifies fluctuations, as we demonstrate below, and is known to restrict the applicability of homogenization approximations in 2D and 3D \citep{Mahiout2020Homogenization}.

The present study develops an approach initiated by \citet{Russell2020Homogenization}, who used an iterative method to approximate the effects of disorder in a linear transport problem involving advection, diffusion and solute uptake via first-order kinetics. They considered a spatially 1D problem with uptake taking place at isolated point sinks.  They considered parameter ranges for which a steady concentration field can be constructed via a smooth (homogenized) leading-order solution, to which corrections are added that account for the discreteness and disorder of the sink distribution.  Corrections are non-local and were evaluated using a Green's function, sidestepping the assumption of unit-cell periodicity that underlies traditional two-scale homogenization. \citet{Russell2020Homogenization} considered a parameter regime in which diffusion was dominant at the intersink distance $\rho$, allowing the use of Riemann sums to approximate certain sums as integrals.  Their approach was constructive: rather than seeking to prove formal convergence, explicit evaluation of the magnitude of corrections allowed domains of validity to be established, and simulation was used to evaluate accuracy.  \chg{ Russell \& Jensen \cite{Russell2020Homogenization} demonstrated improved accuracy of corrections to a leading-order homogenization solution evaluated using a Green's function approach in comparison to classical two-scale asymptotics assuming microscale periodicity.  They also compared the magnitude of corrections to solute fields for periodic, normally-perturbed and uniformly-random sink distributions, each showing distinct dependence on the underlying physical parameters.}

Here we extend this work in four directions, while adopting the same constructive approach: (i) the problem is reformulated to focus on the mapping from statistical moments of the sink distribution to statistical moments of the solute distribution, allowing sink distributions to be represented (for example) as a Gaussian process; (ii) a parameter regime is considered for which advection dominates diffusion over intersink lengthscales, leading to non-smooth concentration profiles; (iii) the study is extended to 2D and 3D, for which the point-sink approximation must be relaxed to allow sinks to have finite size, so that fluctuations remain bounded; (iv) although corrections to a naive homogenization approximation are generally non-local, we show that an essentially local correction to the mean concentration field can be identified when the sink correlation length is sufficient small, and we evaluate this correction explicitly for  sinks distributed uniformly randomly in a 2D or 3D domain. 

\begin{figure}[t!]
\centering
{\includegraphics[width=.8\linewidth]{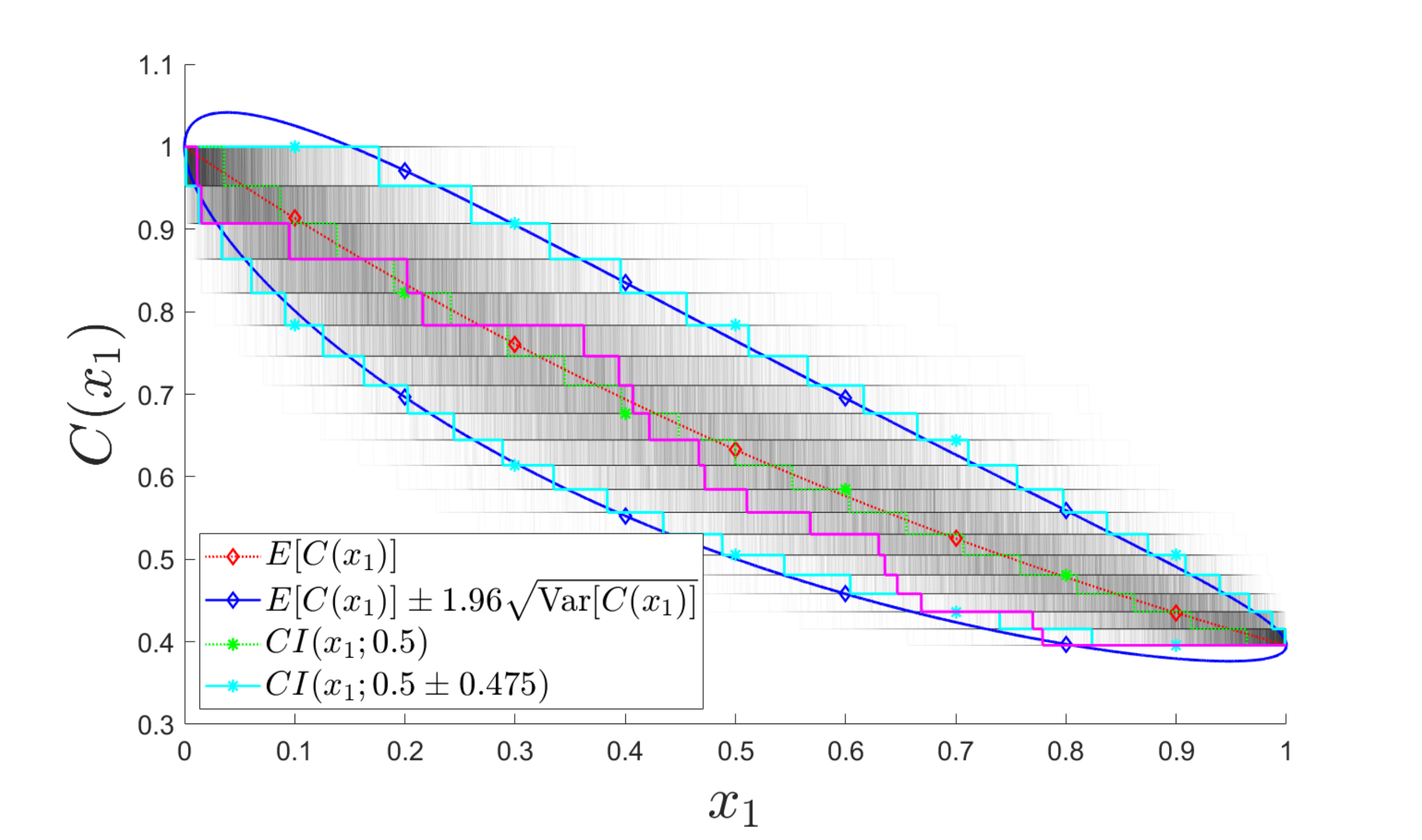}}
\caption{$19$ point sinks are distributed uniformly randomly along the unit interval, with concentration $C(x)$ falling by a factor $1/(1+S_1)$ at each one, where $S_1=0.05$.  From $10^4$ realisations of this process, we show: a single realisation (solid magenta); the full ensemble of $10^4$ concentration profiles (gray); their expectation ($\mathbb{E}[C(x)]$, \ref{ExpZeroDiffusionUniform}, dashed red); Gaussian-based 95\% credible intervals ($\mathbb{E}[C(x)] \pm 1.96 \sqrt{\Var[C(x)]}$, solid blue, using (\ref{VarZeroDiffusionUniform})); median ($CI(x;0.5)$, dashed green, using (\ref{ZeroDiffusionCIUniform}) with $r=1/2$); cdf 95\% credible intervals ($CI(x;0.5\pm0.475)$, solid cyan, \ref{ZeroDiffusionCIUniform}).} \label{fig:RealisationsZeroDiffusion}
\end{figure}

To set the scene, Figure~\ref{fig:RealisationsZeroDiffusion} shows a set of realizations of a 1D advection-uptake process (with no solute diffusion).  In this example, 19 point sinks are distributed randomly in the domain $(0,1)$, each removing a fixed proportion of the oncoming concentration (which takes the value 1 at the inlet at $x=0$ and is swept uniformly in the positive $x$ direction).  An individual realisation (magenta) reveals the staircase structure of a typical 1D concentration field and shows how it deviates appreciably from the discontinuous sample median (green) and the smooth sample mean (red).  This example illustrates how the concentration distribution can be non-Gaussian, with credible intervals (cyan) deviating from the equivalent intervals defined by the sample variance (blue) near the source (where concentrations cannot exceed unity) and near the sink (where concentrations cannot \chg{fall} below $1.05^{-19}\approx 0.396$).  This example illustrates how averaging leads to non-smooth concentration fields having smooth statistical moments, even if these must be interpreted cautiously in some circumstances.  Expressions for the moments and credible intervals of this simple example are derived in Appendix~\ref{app:A}.

While it is relatively straightforward to make use of an exact Green's function for a 1D transport problem (satisfying appropriate inlet and outlet boundary conditions), this is less true in 2D and 3D, and the high-dimensional integrals needed to evaluate higher moments quickly become computationally costly. However when advection dominates diffusion, the free-space Green's function provides a potentially useful simplification.  The Green's function for advection/diffusion/uptake has a discontinuity in 1D, a $\log r$ singularity in 2D and a $1/r$ singularity in 3D, making homogenization feasible for point sinks in 1D \citep{Panasenko2016Homogenization} but more challenging in higher dimensions \citep{Mahiout2020Homogenization}.  Accordingly, we consider below isolated sinks of finite width $\varsigma$, taking them to be distributed uniformly randomly in space.  We formulate a transport problem in a domain that is bounded in the advective direction $x_1$, assuming spatially uniform inlet flux at $x_1=0$, and assume that sink distributions are statistically uniform over a  region that is bounded in the transverse direction.  Despite individual realisations having a complex spatial structure, moments typically depend on $x_1$ alone, and become smooth as a result of averaging.  In the present study we assume that advection is uniform, ignoring heterogeneity of the flow field or of diffusivity, allowing us to exploit a tractable free-space Green's function.

In order to capture the effect of disorder within a homogenization approximation, we also adopt a device described by \citet{Noetinger2018Effective} and exploit the limit in which the correlation length of the covariance of the sink distribution is very small.  In the present example, we show that this length is provided by the sink size $\varsigma$ for sinks distributed uniformly randomly in 2D or 3D.  This allows us to evaluate an effective uptake parameter $\Da_{\eff}$: replacing the dimensionless Damk\"ohler number in the naive homogenized solution with $\Da_{\eff}$, we obtain a direct approximation for the mean concentration that quantifies how disorder reduces uptake when sinks are distributed uniformly randomly in 2D or 3D.  

The model that we investigate is outlined in Section~\ref{sec:2}\ref{sec:Model}, with example simulations presented in Section~\ref{sec:2}\ref{sec:2dsim}.  The moments-based expansion is presented in Section~\ref{sec:2}\ref{sec:DevelopingApproach}, revealing the critical roles of the Green's function (Section~\ref{sec:2}\ref{sec:free}) and its singularities in the evaluation of high-dimensional integrals (Section~\ref{sec:2}\ref{sec:inteval}).  The derivation of $\Da_{\mathrm{eff}}$ is given in Section~\ref{sec:2}\ref{sec:daeff}.  Predictions are evaluated against simulations in Section~\ref{sec:3}. 

\section{Model and Methods}
\label{sec:2}

\subsection{The model problem}
\label{sec:Model}

We formulate the model in 3D, adopting analogues in 1D and 2D when required. Let $\mathcal{D}_3$ be a domain of thickness $L$ defined such that $ \mathbf{x}^* = (x_1^*, x_2^*, x_3^*) \in \mathcal{D}_3$ when $x_1^* \in [0, L]$ and $x_2^*, x_3^* \in \mathbb{R}$. $C^*(\mathbf{x}^*;\omega)$, $U$, $D$ and $S$ represent the \chg{(dimensional)} solute concentration field, uniform advective velocity in the $x_1^*$ direction, diffusion coefficient and uptake rate respectively. 
Uptake is mediated by a distributed sink function satisfying $1 + \hat{g}^*(\mathbf{x}^*;\omega) \ge 0$, \chg{where $\hat{g}^*$ has zero spatial average}. $\omega$ denotes that $\hat{g}^*(\mathbf{x}^*;\omega)$ is a realisation drawn from a prescribed distribution, making $C^*(\mathbf{x}^*;\omega)$ a random variable.  

We prescribe a solute flux $q$ on the plane $x_1^*=0$, with zero diffusive flux on $x_1^*=L$ and as $x_2^*, x_3^* \rightarrow \pm \infty$. Defining $ \mathbf{x} = \mathbf{x}^*/L$, $\hat{g}(\mathbf{x};\omega) = \hat{g}^*(\mathbf{x}^*;\omega)$ and $C(\mathbf{x};\omega) = C^*(\mathbf{x}^*;\omega)/(q/U_0)$, the dimensionless concentration satisfies the advection-diffusion-uptake equation 
\begin{subequations}\label{GoverningEquation}
\begin{gather}
\nabla^2_{3D} C - \Pe \partial_{x_1} C - \Da C (1+\hat{g}(\mathbf{x};\omega)) = 0
\end{gather}
and boundary conditions
\begin{gather}
(1 - \Pe^{\,-1} \partial_{x_1})C|_{x_1=0} =  1, \quad \partial_{x_1}C|_{x_1=1} = 0, \quad \partial_{x_2}C|_{x_2\rightarrow \pm \infty} \rightarrow 0, \quad \partial_{x_3}C|_{x_3 \rightarrow \pm \infty} \rightarrow 0,
\end{gather}
\end{subequations}
where $x_1 \in [0, 1]$, $x_2, x_3\in \mathbb{R}$ and $\nabla^2_{3D} \equiv  \partial_{x_1^2} + \partial_{x_2^2} + \partial_{x_3^2}$.
The P\'eclet number $\Pe = U L / D$ represents the strength of advection to diffusion; the Damk\"ohler number $\Da = S{L^2}/D$ relates the rate of uptake to diffusion. We focus here on the strong-advection regime $\Pe\gg \max(1,\sqrt{\mathrm{Da}})$; of particular interest is the distinguished limit in which $\Pe/\Da=U/S{L}=O(1)$, implying a balance between advection and uptake across the whole domain.

Isolated sinks are taken to be of finite size and to occupy a subdomain $\mathcal{D}_3^s$ of $\mathcal{D}_3$ in which $x_1 \in [0, 1]$ and $x_2, x_3 \in [-L_s, L_s] $. Let $\rho = 1/N$ be the average inter-sink distance in any direction, 
where $N \in \mathbb{Z}^+$ represents the number of sinks per unit length. Let the midpoint of sink locations be represented by $\bm{\xi}_{\mathbf{i}_3} = (\xi_i,\xi_j,\xi_k)$, where $\mathbf{i}_3 \in \{i,j,k\}$, $i = 1,\dots,N$ and $j,k = -M, \dots, M $ with $M=\lfloor L_s N \rfloor \in \mathbb{Z}$. Thus there are $(2M+1)^2/\rho$ sinks in $\mathcal{D}_3^s$ with an average density per unit volume given by $\rho^{-3}$. We define $\hat{g}(\mathbf{x};\omega)$ to be
\begin{equation}\label{g-discrete}
\hat{g}(\mathbf{x};\omega) = \rho^3 \textstyle{\sum_{\mathbf{i}_3}} F^{(3)}_{\varsigma}(\mathbf{x} - \bm{\xi}_{\mathbf{i}_3}) - 1,
\end{equation}
where $\sum_{\mathbf{i}_3} \equiv \sum_{i=1}^{N}\sum_{j=-M}^{M}\sum_{k=-M}^{M}$ and $F^{(3)}_{\varsigma}(\mathbf{x} - \bm{\xi}_{\mathbf{i}_3})$ is a regularised uptake function with width $\varsigma \ll 1$ such that
\begin{equation}\label{requirement}
\int_{\mathcal{D}_3^s} F^{(3)}_{\varsigma}(\mathbf{x} - \bm{\xi}_{\mathbf{i}_3}) \, \mathrm{d} \bm{\xi}_{\mathbf{i}_3} = 1.
\end{equation}
This choice of $F^{(3)}_{\varsigma}$ ensures $\hat{g}(\mathbf{x};\omega)$ has a spatially-averaged density of zero within $\mathcal{D}_3^s$.  We assume throughout that isolated sinks have multivariate uniform distribution, 
such that $\xi_i \sim \mathcal{U}[0,1]$ and $\xi_j, \xi_k \sim \mathcal{U}[-L_s,L_s]$.  Similar definitions of the sink function can be made for a 1D [2D] domain $\mathcal{D}_1$ [$\mathcal{D}_2$], where $F^{(3)}_{\varsigma}$ is replaced by $F^{(1)}_{\varsigma}$ [$F^{(2)}_{\varsigma}$], volumes ($\rho^3$) are replaced by distances ($\rho$) [areas ($\rho^2$)] and triple-sums over $\mathbf{i}_3 \in \{i,j,k\}$ are replaced by single- [double-] sums over $\mathbf{i}_1 = i$ [$\mathbf{i}_2 \in \{i,j\}$]. We adopt the Gaussian sink structure function
\begin{equation}\label{F_varsigma}
F^{\chg{(}n\chg{)}}_{\varsigma}(\mathbf{x} - \mathbf{x}_{\mathbf{i}_n}) = \dfrac{1}{(2\pi\varsigma^2)^{n/2}}\exp \left( - \dfrac{1}{2 \varsigma^2}|\mathbf{x} - \mathbf{x}_{\mathbf{i}_n}|^2 \right),
\end{equation}
where $\varsigma$ remains sufficiently small to satisfy (\ref{requirement}) and prevent sinks from overlapping, to exponential accuracy. \chg{This function is chosen for convenience but could be replaced to model specific applications.}

It will be helpful to represent distributions of isolated sinks in terms of their first two statistical moments.  As shown in Appendix~\ref{sec:SinkFunctionMoments}, uniformly-random sinks with Gaussian structure function (\ref{F_varsigma}) have ensemble mean and covariance \begin{equation}\label{Cov_g_UR_n_extra}
\mathbb{E}[\hat{g}]=0, \quad 
\mathcal{K}_{\hat{g}}[\mathbf{x},\mathbf{y}] = \rho^n F^{(n)}_{\sqrt{2}\varsigma}(\mathbf{x} - \mathbf{y}) - \dfrac{\rho}{(2M+1)^{n-1}},
\end{equation}
where $\mathcal{K}_{f}[\mathbf{x},\mathbf{y}] \equiv \mathcal{K}[f(\mathbf{x};\omega),f(\mathbf{y};\omega)] $ and $\mathcal{K}$ represents covariance. An important distinction between 1D and higher-dimensional cases is evident.  For $n=1$, $\mathcal{K}_{\hat{g}}$ has a non-local contribution (with $N$ sinks in a 1D domain, finding one sink at a location reduces slightly the chance of finding another elsewhere).  However for $n>1$, with $M\rightarrow \infty$, the nonlocal term vanishes (because the sinks \chg{can} occupy an arbitrarily wide area or volume \chg{within $\mathcal{D}_2$ or $\mathcal{D}_3$}).  The sink density in this case resembles a Gaussian process with square-exponential covariance  $\sigma^2 \exp(-\vert\mathbf{x}-\mathbf{y}\vert^2/\ell^2)$, having variance and correlation length given respectively by
\begin{equation}
\label{eq:5a}
    \sigma^2=(\rho/(2\sqrt{\pi}\varsigma))^n, \quad \ell = 2\varsigma.
\end{equation}

\subsection{2D simulations}
\label{sec:2dsim}

\begin{figure}[t!]
\centering
\subfloat[]{
\raisebox{-9.28cm}{
\begin{tikzpicture}\draw (0,0) node[above   right]{\includegraphics[height=.5\linewidth]{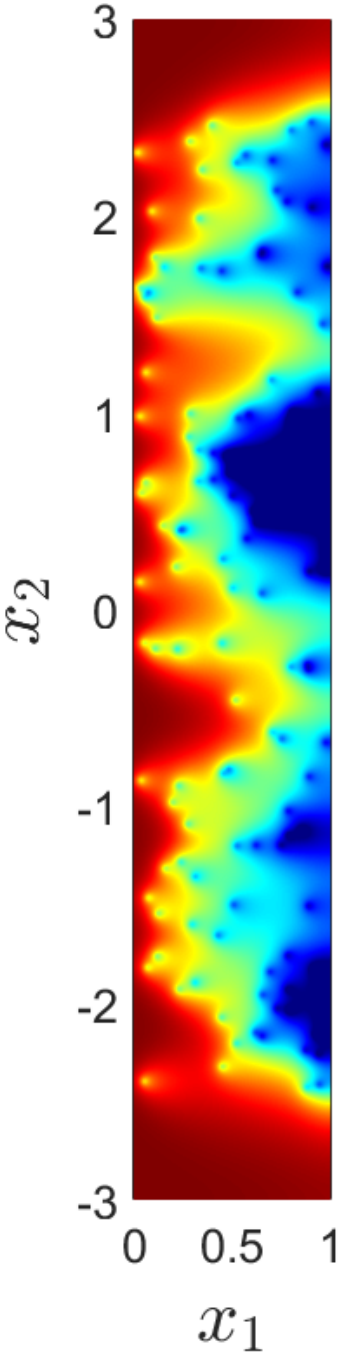}};
\draw (2.5,0.93) node[above   right]{\includegraphics[height=.445\linewidth]{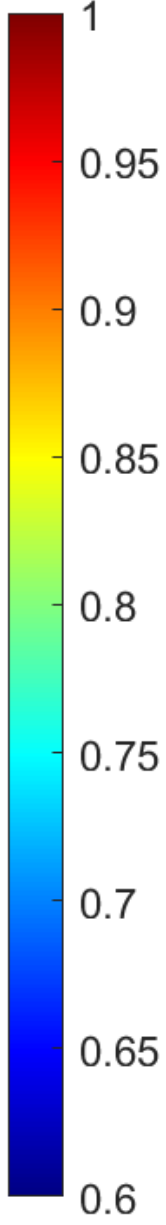}};
\end{tikzpicture}}} \hspace{0.3cm}
\subfloat[]{
\raisebox{-9.28cm}{
\begin{tikzpicture}
\draw (0,0) node[above
right]{\includegraphics[height=.5\linewidth]{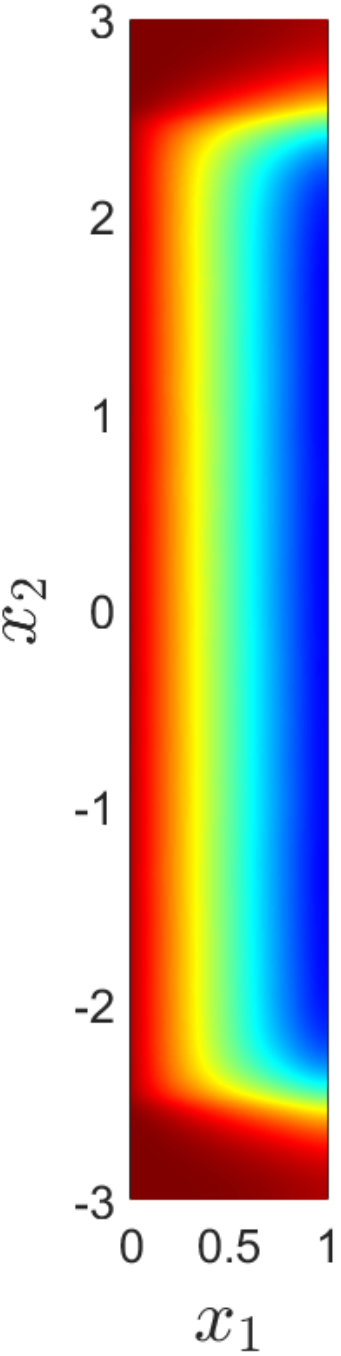}};
\draw (2.5,0.93) node[above   right]{\includegraphics[height=.445\linewidth]{colorbar.pdf}};
\end{tikzpicture}}} \hspace{0.3cm}
\subfloat[]{
\raisebox{-9.28cm}{
\begin{tikzpicture}
\draw (8,0) node[above right]{\includegraphics[height=.52\linewidth]{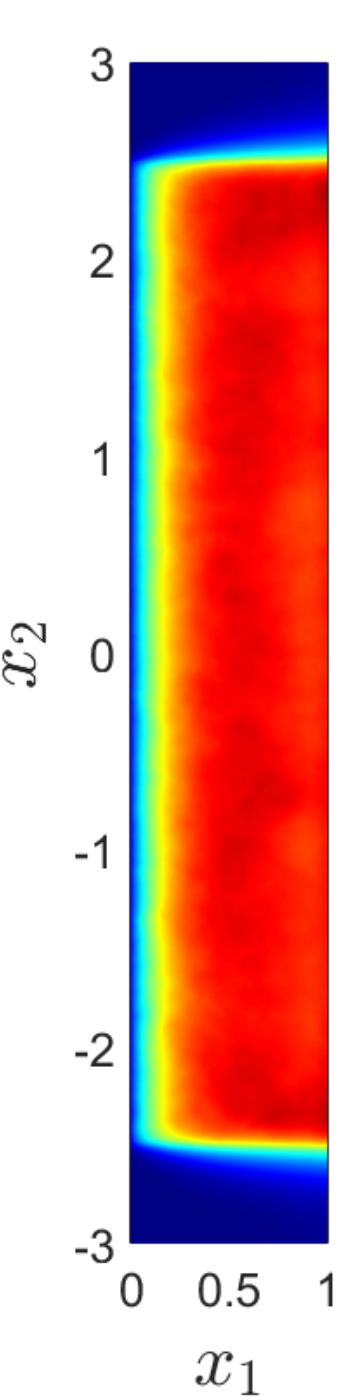}};
\draw (10.55,0.93) node[above   right]{\includegraphics[height=.47\linewidth]{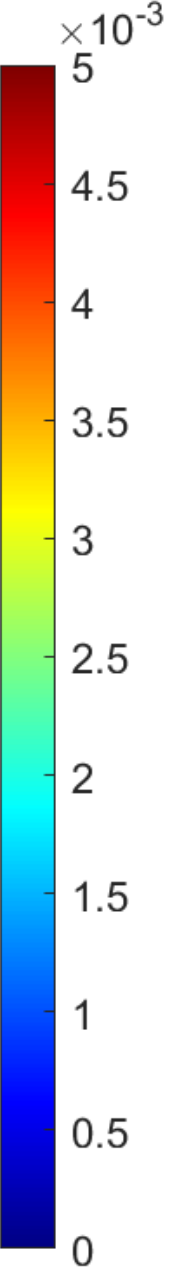}};
\end{tikzpicture}}}
\caption{2D solute concentration satisfying (\ref{GoverningEquation}) for sinks located uniformly randomly in the domain $\mathcal{D}_2^s = [0, 1] \times [-2.5,2.5]$ for $\rho=0.2$, $(\Pe, \Da) = (20, 10)$ and $\varsigma=0.01$: (a)  a single realisation; (b) sample expectation and (c) sample variance, calculated from $10^4$ realisations.}
\label{fig:Realisations_2D_L3}
\end{figure}

Realisations of concentration fields were calculated numerically using a second-order-accurate finite-difference scheme.  Representative simulations in 2D are shown in Figure \ref{fig:Realisations_2D_L3}.  While an individual realisation shows strong disorder, with clear evidence of left-to-right advection (Figure~\ref{fig:Realisations_2D_L3}a), the mean concentration field and its variance become smooth and independent of $x_2$ when sufficiently far from the boundaries of $\mathcal{D}_2^s$ at $x_2=\pm 2.5$ (Figure~\ref{fig:Realisations_2D_L3}b,c).  This arises through a combination of averaging effects and strong advection, which limits the degree of lateral diffusive spread downstream of each sink. We seek approximations of these smooth 1D functions in terms of the sink density $\rho$, sink width $\varsigma$ and the physical parameters $\Pe$ and $\Da$.

\subsection{A moments-based expansion}\label{sec:DevelopingApproach}

The volume-averaged sink density in $\mathcal{D}_3^s$ is unity, making it natural to define the \chg{leading-order} homogenized linear and boundary operators associated with (\ref{GoverningEquation}) as $\mathcal{L}_3 \equiv \nabla^2_{3D} - \Pe\partial_{x_1} - \Da$ and
\begin{multline*}
\mathcal{B}_3 = \{ \left( 1 - (1/\Pe)\partial_{x_1} \right)(\cdot)|_{x_1=0} , \, \partial_{x_1}(\cdot) |_{x_1=1}, \,\partial_{x_2}(\cdot) |_{x_2 \rightarrow -\infty}, \partial_{x_2}(\cdot) |_{x_2 \rightarrow \infty}, 
 \\ 
\partial_{x_3}(\cdot) |_{x_3 \rightarrow -\infty}, \, \partial_{x_3}(\cdot) |_{x_3 \rightarrow \infty} \}
\end{multline*}
respectively. The \chg{leading-order} homogenized solution $C_H(\mathbf{x})$ associated with (\ref{GoverningEquation}) can be found by solving
\begin{equation}\label{C_H_eqn}
\mathcal{L}_3 C_H(\mathbf{x}) = 0, \quad \mathcal{B}_3 C_H(\mathbf{x}) = \{ 1 , 0, 0, 0, 0, 0 \}.
\end{equation}
It is evident that 
$C_H(\mathbf{x})$ depends only on $x_1$, being
\begin{equation}\label{C_H}
C_H(x_1) = \dfrac{\Pe}{\psi(1)} \Big( (2\phi - \Pe)e^{\phi(x_1-1)} + (2\phi + \Pe) e^{\phi(1-x_1)} \Big)e^{(\Pe/2)x_1},
\end{equation}
where $\phi \equiv \sqrt{\Pe^{\,2}/4 + \Da}$ and $\psi(x_1) \equiv (2\Pe\phi + \Pe^{\,2} + 2\Da)e^{\phi x_1} + (2\Pe\phi - \Pe^{\,2} - 2\Da)e^{-\phi x_1}$. 
In the limit $\Pe\gg \max(1,\sqrt{\Da})$ of interest here, $C_H\approx \exp[-\Da x_1/\Pe]$, showing how the concentration decays over a lengthscale defined by a balance between uptake and advection.  Writing the concentration as
\begin{equation}\label{expansion}
C(\mathbf{x};\omega) = C_H(x_1) 
+ \Da \widehat{C}_1(\mathbf{x}; \omega) + \Da^2 \widehat{C}_2(\mathbf{x}; \omega) + \dots,
\end{equation}
we construct a solution of (\ref{GoverningEquation}), to be validated \textit{a posteriori}, using the ansatz
\begin{subequations}\label{LinearOperatorEqns}
\begin{gather}
\mathcal{L}_3\widehat{C}_1(\mathbf{x};\omega) = \hat{g}(\mathbf{x};\omega) C_H(x_1), \quad \mathcal{B}_3\widehat{C}_1(\mathbf{x}; \omega) = \{0, \dots, 0\}, \\
\mathcal{L}_3\widehat{C}_2(\mathbf{x};\omega) = \hat{g}(\mathbf{x};\omega) \widehat{C}_1(\mathbf{x};\omega), \quad \mathcal{B}_3\widehat{C}_2(\mathbf{x}; \omega) = \{0, \dots, 0\}, 
\end{gather}
\end{subequations}
etc.  To invert the linear operators in (\ref{LinearOperatorEqns}), we define $G_3(\mathbf{x},\mathbf{x}')$ to be the associated 3D Green's function satisfying
\begin{equation} \label{G_equation}
\mathcal{L}_3G_3(\mathbf{x},\mathbf{x}') = \delta(\mathbf{x}-\mathbf{x}'), \quad \text{ where } \quad \mathcal{B}_3G_3(\mathbf{x},\mathbf{x}') = \{0, \dots, 0\}.
\end{equation}
Applying homogeneous boundary conditions in the $x_2$- and $x_3$-directions is appropriate as the source term is compact. The Green's function can then be used to give the corrections
\begin{subequations}
\begin{align}
\widehat{C}_1(\mathbf{x};\omega) &= \int_{\mathcal{D}_3} G_3(\mathbf{x}, \mathbf{x}') C_H(x_1') \hat{g}(\mathbf{x}';\omega) \, \mathrm{d}\mathbf{x}', \label{correction1} \\
\widehat{C}_2(\mathbf{x};\omega) &= \int_{\mathcal{D}_3} \int_{\mathcal{D}_3} G_3(\mathbf{x}, \mathbf{x}') G_3(\mathbf{x}', \mathbf{x}'') C_H(x_1'') \hat{g}(\mathbf{x}';\omega) \hat{g}(\mathbf{x}'';\omega) \, \mathrm{d}\mathbf{x}' \, \mathrm{d}\mathbf{x}''. \label{correction2}
\end{align}
\end{subequations}
We characterise the corrections in terms of their moments evaluated over realisations, specifically
\begin{subequations}
\label{eq:moments}
\begin{align}
\mathbb{E}\left[\widehat{C}_1(\mathbf{x};\omega)\right] & = \int_{\mathcal{D}_3} G_3(\mathbf{x},\mathbf{x}') C_H(x_1') \mathbb{E} [\hat{g}(\mathbf{x}';\omega)] \, \mathrm{d}\mathbf{x}', \label{ExpC1}\\
\mathcal{K}_{\widehat{C}_1}[\mathbf{x},\mathbf{y}] & = \int_{\mathcal{D}_3} \int_{\mathcal{D}_3} G_3(\mathbf{x},\mathbf{x}') C_H(x_1') \mathcal{K}_{\hat{g}}[\mathbf{x}',\mathbf{y}'] G_3(\mathbf{y},\mathbf{y}') C_H(y_1') \, \mathrm{d}\mathbf{x}' \, \mathrm{d}\mathbf{y}', \label{CovC1} \\
\mathbb{E}\left[\widehat{C}_2(\mathbf{x};\omega)\right] & = \int_{\mathcal{D}_3} \int_{\mathcal{D}_3} G_3(\mathbf{x},\mathbf{x}') G_3(\mathbf{x}',\mathbf{x}'') C_H(x_1'') \mathbb{E}\left[\hat{g}(\mathbf{x}';\omega) \hat{g}(\mathbf{x}'';\omega)\right] \, \mathrm{d}\mathbf{x}' \, \mathrm{d}\mathbf{x}''. \label{ExpC2}
\end{align}
\end{subequations}
This approach extends to $n=1,2$ dimensions, replacing $\mathcal{D}_3$ and $G_3(\mathbf{x},\mathbf{x}')$ with $\mathcal{D}_n$ and $G_n(\mathbf{x},\mathbf{x}')$ respectively, generalising the 1D formulation in \citet{Russell2020Homogenization}.  In higher dimensions, complications emerge due to singularities of $G_2$ and $G_3$ as $\mathbf{x} \rightarrow \mathbf{x}'$ and the high dimensionality of the quadrature.

\subsection{The free-space Green's function}
\label{sec:free}

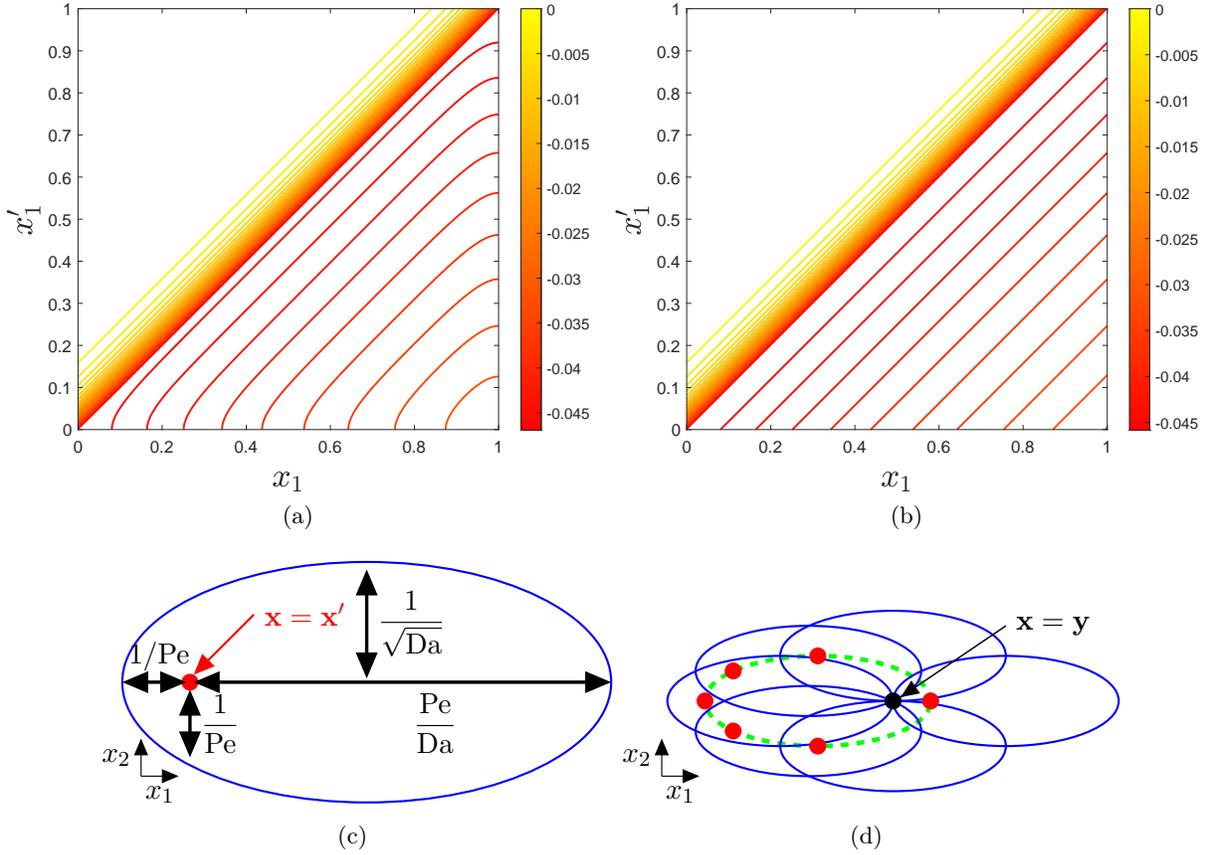
\begin{figure}[t!]
\centering
\subfloat[]{\includegraphics[height=.35\linewidth]{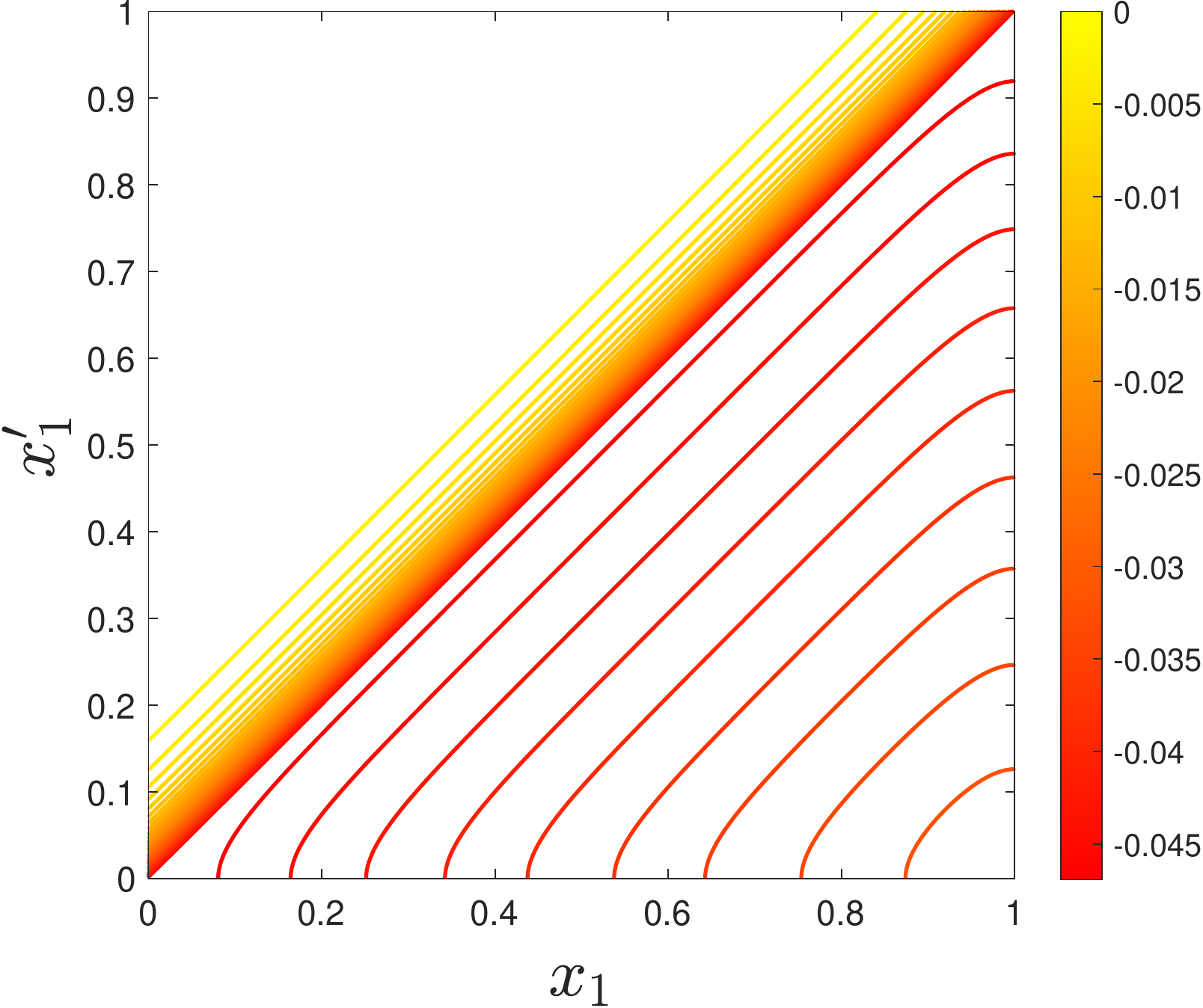}}
\hspace{0.2cm}
\subfloat[]{\includegraphics[height=.35\linewidth]{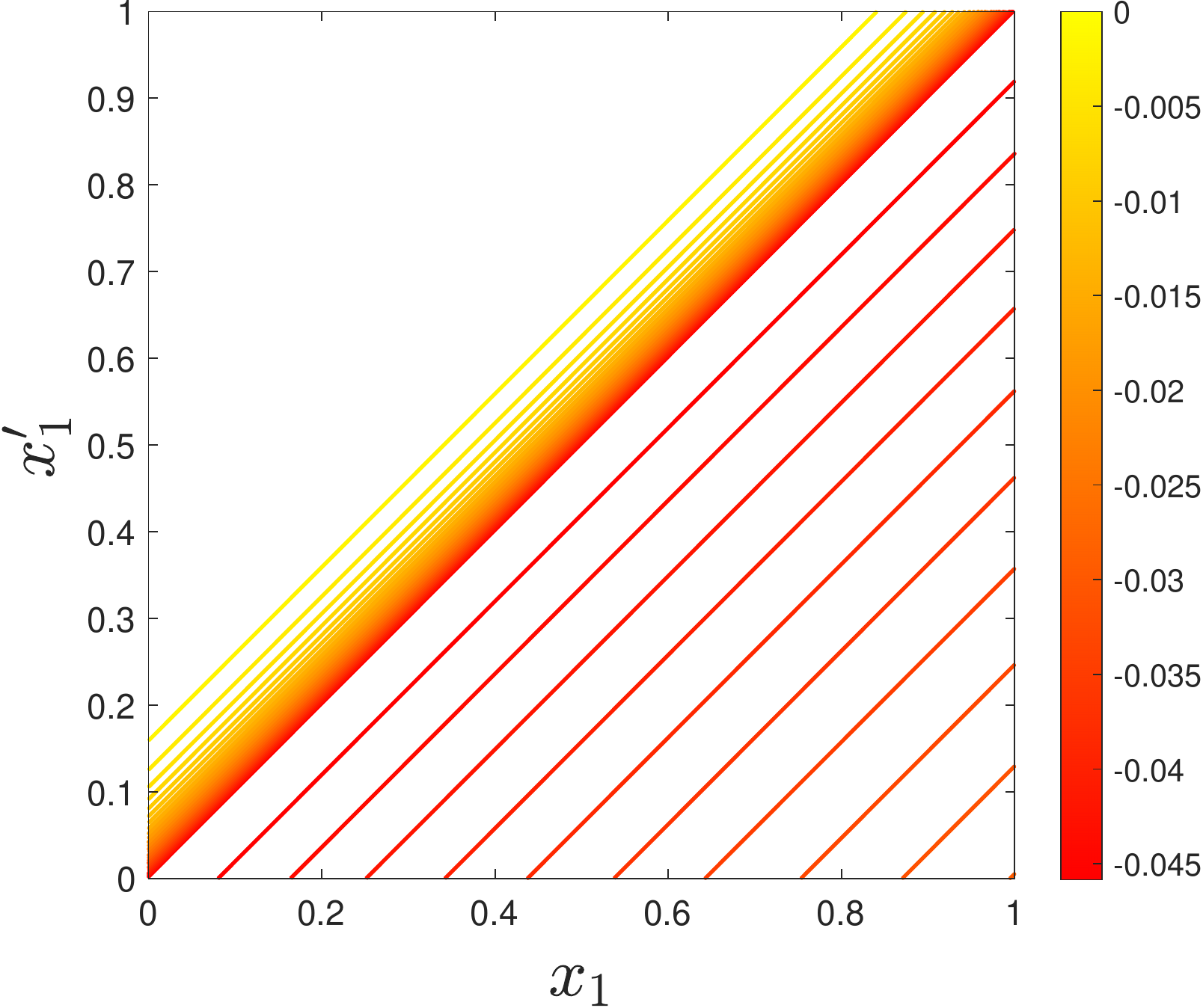}} \hspace{0.1cm}

\subfloat[]{\begin{tikzpicture}
\draw[blue,thick] (0,0) ellipse (3.25cm and 1.6cm);

\draw[>=triangle 45, <-,line width=0.2mm, red, thick] (-2.3,0.1) -- (-1.5,0.9) node[above, right]{\color{red}{$\mathbf{x}=\mathbf{x}'$}};
\filldraw[red] (-2.35,0) circle (3pt);

\draw[>=triangle 45, <->,line width=0.45mm] (-3.25,0) -- (-2.4,0) node[midway,above]{~${1}/{\Pe}$};
\draw[>=triangle 45, <->,line width=0.45mm] (-2.3,0) -- (3.25,0) node[midway,below right]{$\dfrac{\Pe}{\Da}$};
\draw[>=triangle 45, <->,line width=0.45mm] (-2.35,-0.1) -- (-2.35,-1) node[midway,right]{$\dfrac{1}{\Pe}$};
\draw[>=triangle 45, <->,line width=0.45mm] (0,0.05) -- (0,1.5) node[midway,right]{$\dfrac{1}{\sqrt{\Da}}$};

\draw[>=triangle 45,->, line width=0.15mm] (-3,-1.25) -- (-3,-0.75) node[midway, left]{$x_2$};
\draw[>=triangle 45,->, line width=0.15mm] (-3,-1.25) -- (-2.5,-1.25) node[midway, below]{$x_1$};
\end{tikzpicture}}
\subfloat[]{\begin{tikzpicture}
\draw[dashed,green,line width=0.6mm] (-2.5,0) ellipse (1.5cm and 0.6cm);

\draw[blue,thick] (0,0) ellipse (1.5cm and 0.6cm);
\draw[blue,thick] (-1.5,0.6) ellipse (1.5cm and 0.6cm);
\draw[blue,thick] (-1.5,-0.6) ellipse (1.5cm and 0.6cm);
\draw[blue,thick] (-2.625,0.4) ellipse (1.5cm and 0.6cm);
\draw[blue,thick] (-2.625,-0.4) ellipse (1.5cm and 0.6cm);
\draw[blue,thick] (-3,0) ellipse (1.5cm and 0.6cm);

\filldraw[red] (-1,0) circle (3pt);
\filldraw[red] (-2.5,0.6) circle (3pt);
\filldraw[red] (-2.5,-0.6) circle (3pt);
\filldraw[red] (-3.625,0.4) circle (3pt);
\filldraw[red] (-3.625,-0.4) circle (3pt);
\filldraw[red] (-4,0) circle (3pt);
\filldraw (-1.5,0) circle (3pt);
\draw[>=triangle 45, <-,line width=0.2mm] (-1.4,0.05) -- (0,1) node[above, right]{$\mathbf{x}=\mathbf{y}$};

\draw[>=triangle 45,->, line width=0.15mm] (-4.575,-1) -- (-4.575,-0.5) node[midway, left]{$x_2$};
\draw[>=triangle 45,->, line width=0.15mm] (-4.575,-1) -- (-4.075,-1) node[midway, below]{$x_1$};
\end{tikzpicture}}
\caption{(a) Exact ${G}_1(x_1,x_1')$ and (b) free-space $\mathcal{G}_1(x_1-x_1')$ Green's function  in 1D, given by (\ref{G_1D}) and (\ref{Free-Space-GF-1D})  respectively, for $(\Pe, \Da) = (20, 10)$.  
(c) Sketch of lengthscales involved in the 2D Green's function for a sink located at $\mathbf{x}=\mathbf{x}'$ [red dot] and the asymptotic shape of the wake [solid blue], for $\Pe \gg \max(1,\sqrt{\Da})$. (d) The \chg{asymptotic} region of influence [dashed green] about the point $\mathbf{x}=\mathbf{y}$ [black dot]. Sinks located outside of this region will have \chg{significantly weaker} influence on the concentration at $\mathbf{x}=\mathbf{y}$ \chg{than those inside}. Red dots represent sink locations $\mathbf{x}=\mathbf{x}'$ and blue ellipses represent the asymptotic shapes of the wake about each sink.}\label{fig:GreensFunctionlengthscales}
\end{figure}

While the Green's function in 1D is straightforward to evaluate (Appendix \ref{app:green}), it is convenient to instead use the free-space Green's function  $\mathcal{G}_n(\mathbf{x} -\mathbf{x}')$ for computations in higher dimensions.  In 3D, this satisfies $\mathcal{L}_3\mathcal{G}_3(\mathbf{x} -\mathbf{x}') = \delta(\mathbf{x}-\mathbf{x}')$ and $\mathcal{G}_3(\mathbf{x})\rightarrow 0$ as $\vert \mathbf{x}\vert\rightarrow \infty$.  $\mathcal{G}_n$ is given by (\ref{Free-Space-GF}): it shares with $G_n$ the $\log(\phi|\mathbf{x} - \mathbf{x}'|)$ singularity in 2D and $1/|\mathbf{x} - \mathbf{x}'|$ singularity in 3D.  $\mathcal{G}_n$ offers a close approximation of $G_n$ in the limit $\Pe \gg \max(1,\sqrt{\Da})$, as illustrated for $n=1$ in  Figure~\ref{fig:GreensFunctionlengthscales}(a,b).  This shows a discrepancy between $G_1(x_1, x_1')$ and $\mathcal{G}_1(x_1-x_1')$ only within a $1/\Pe$ distance of the outlet in $x_1$ and the inlet in $x_1'$.  The identity
\begin{equation}\label{GF-relation}
\int_{-\infty}^{\infty} \int_{-\infty}^{\infty} \mathcal{G}_3(\mathbf{x}) \, \mathrm{d}x_2 \, \mathrm{d}x_3 = \int_{-\infty}^{\infty} \mathcal{G}_2(\mathbf{x}) \, \mathrm{d}x_2 = \mathcal{G}_1(x_1).
\end{equation}
 will allow us to make use of $\mathcal{G}_1$ later on.

$\mathcal{G}(\mathbf{x}-\mathbf{x}')$ denotes the field in the $\mathbf{x}$ plane generated by a point sink at $\mathbf{x}'$.  In 2D [3D], concentration contours have an approximately elliptical [ellipsoidal] shape, with dimensions illustrated in Figure~\ref{fig:GreensFunctionlengthscales}(c) when $\Pe\gg\max(1,\sqrt{\Da})$, as explained in Appendix~\ref{app:green}.  We can use this structure to identify the \chg{asymptotic} region of influence associated with a point $\mathbf{x}$, within which sources at $\mathbf{x}'$ will contribute \chg{appreciably} to the concentration field at $\mathbf{x}$, as illustrated in Figure~\ref{fig:GreensFunctionlengthscales}(d).  Strong advection implies that the region of influence is largely upstream of $\mathbf{x}$, while strong uptake ensures that the region is narrow in the direction transverse to the flow. This allows quadrature to be restricted to physically relevant domains.


\subsection{Evaluation of moments}
\label{sec:inteval}

Adopting the free-space Green's function approximation and incorporating the sink moments (\ref{Cov_g_UR_n_extra}), (\ref{eq:moments}) becomes
\begin{subequations}
\label{eq:expc1}
\begin{align}
\label{ExpC1_Uniform}
\mathbb{E}\left[\widehat{C}_1(\mathbf{x};\omega)\right] &= 0,\\
\mathcal{K}_{\widehat{C}_1}[\mathbf{x},\mathbf{y}] &= \int_{\mathcal{D}_n} \int_{\mathcal{D}_n} \mathcal{G}_n(\mathbf{x}-\mathbf{x}') C_H(x_1') \mathcal{G}_n(\mathbf{y}-\mathbf{y}') C_H(y_1') \nonumber
 \\ & \qquad \qquad \qquad \qquad \times
\left(\rho^n F^{(n)}_{\sqrt{2}\varsigma}(\mathbf{x}' - \mathbf{y}') - \dfrac{\rho}{(2M+1)^{n-1}} \right) \, \mathrm{d}\mathbf{x}' \, \mathrm{d}\mathbf{y}',
\label{CovC1_Uniform} \\
\mathbb{E}\left[\widehat{C}_2(\mathbf{x};\omega)\right] & = \int_{\mathcal{D}_n} \int_{\mathcal{D}_n}\mathcal{G}_n(\mathbf{x}-\mathbf{x}') \mathcal{G}_n(\mathbf{x}'-\mathbf{x}'') C_H(x_1'') \nonumber 
\\ & \qquad \qquad \qquad \qquad \times 
\left(\rho^n F^{(n)}_{\sqrt{2}\varsigma}(\mathbf{x}' - \mathbf{x}'') - \dfrac{\rho}{(2M+1)^{n-1}} \right) \, \mathrm{d}\mathbf{x}' \, \mathrm{d}\mathbf{x}''. \label{ExpC2_Uniform}
\end{align}
\end{subequations}
We now consider approximations when the domain width is large ($L_s \gg \rho$) and the sink width small ($\varsigma \rightarrow 0$).  To approximate the variance of $\widehat{C}_1$ in this limit, we can replace $F^{(n)}_{\sqrt{2}\varsigma}$ in (\ref{CovC1_Uniform}) with an $n$-dimensional $\delta$-function  
and note that the second integral in (\ref{CovC1_Uniform}) can be reduced using (\ref{GF-relation}), giving
\begin{multline}\label{VarC1_Uniform_delta}
\Var_{\varsigma \rightarrow 0}[\widehat{C}_1(\mathbf{x},\omega)] = \rho^{n} \int_{\mathcal{D}_n} (\mathcal{G}_n(\mathbf{x}-\mathbf{x}') C_H(x_1'))^2 \, \mathrm{d}\mathbf{x}'  
- \dfrac{\rho}{(2M+1)^{n-1}} \left( \int_{\mathcal{D}_1} \mathcal{G}_1(x_1-x_1') C_H(x_1') \, \mathrm{d}x_1' \right)^2.
\end{multline}
This reduces the $2n$-dimensional integral (\ref{CovC1_Uniform}) to a cheaper $n$-dimensional integral (\ref{VarC1_Uniform_delta}), although some loss of accuracy is anticipated by igorning the finite sink size. 

While (\ref{GF-relation}) can also be used to reduce the second integral in (\ref{ExpC2_Uniform}) to 1D, a $\delta$-function approximation cannot be used for the first integral in $\mathbb{E}[\widehat{C}_2]$ because of singularities in $\mathcal{G}_2$ and $\mathcal{G}_3$.  Instead, we exploit the fact that $F^{(n)}_{\sqrt{2}\varsigma}(\mathbf{x}' - \mathbf{x}'')$ is asymptotically small  when $\varsigma \ll 1$ unless $\mathbf{x}'$ is within an $O(\varsigma)$ distance of $\mathbf{x}''$.  $C_H(x_1'') \approx C_H(x_1')$ over this region while $\mathcal{G}_n (\mathbf{x}' - \mathbf{x}'')$ can be approximated by its leading-order singular form.  We summarise the results of this calculation (see Appendix~\ref{app:Y}),
as $\varsigma \rightarrow 0$ in $n$ dimensions, as
\begin{subequations}
\label{eq:mc2}
\begin{multline} \label{ExpC2_Uniform_simplified}
\mathbb{E}\left[\widehat{C}_2(\mathbf{x};\omega)\right] \approx - \rho^n \beta_n \int_{\mathcal{D}_1} \mathcal{G}_1(x_1 - x_1') C_H(x_1') \, \mathrm{d}x_1' \\
- \dfrac{\rho}{(2M+1)^{n-1}} \int_{\mathcal{D}_1} \int_{\mathcal{D}_1} \mathcal{G}_1(x_1-x_1') \mathcal{G}_1(x_1'-x_1'') C_H(x_1'') \, \mathrm{d}x_1' \, \mathrm{d}x_1'',
\end{multline}
where
\begin{equation}\label{beta_n}
\beta_1 = \dfrac{1}{2 \phi}, \quad \beta_2 = \dfrac{1}{4 \pi} \left(\gamma - 2 \log(2 \phi \varsigma)\right) \quad \text{ and } \quad \beta_3 = \dfrac{1}{4 \pi^{3/2} \varsigma}
\end{equation}
\end{subequations}
and $\gamma$ is the Euler--Mascheroni constant.  The correction in 1D is independent of the sink size $\varsigma$ as $\varsigma \rightarrow 0$, whereas in 2D and 3D the correction grows in magnitude as $\varsigma$ becomes asymptotically small.  In 2D and 3D, when $L_s\gg \rho$, the final terms of $O(\rho/M^{n-1})$ may be neglected and moments become independent of $x_2$ and $x_3$ when suitably far from boundaries, as illustrated in Figure~\ref{fig:Realisations_2D_L3}(b,c). 

Having replaced the exact Green's function by its free-space form, a further approximation can be obtained by neglecting boundary layers of thickness $O(1/\Pe)$ upstream of sinks, evident in Figure \ref{fig:GreensFunctionlengthscales}.  In 1D, we adopt the leading-order expressions 
$C_H\approx \mathrm{e}^{-(\Da/\Pe) x_1}$, $\mathcal{G}_1(x_1-x_1')\approx -({1}/{\Pe}) \mathrm{e}^{-(\Da/\Pe)(x_1-x_1')}H(x_1-x_1')$ for $\Pe\gg 1$, accounting only for the downstream influence of one sink on another. Direct evaluation of (\ref{VarC1_Uniform_delta}) and (\ref{ExpC2_Uniform_simplified}) gives
\begin{equation}
\label{eq:1dvarapprox}
    \mathrm{Var}[\widehat{C}_1(x_1,\omega)]\approx  \frac{\rho}{\Pe^2} (x_1-x_1^2)\mathrm{e}^{-({2\Da}/{\Pe})x_1}, \quad
    \mathbb{E}[\widehat{C}_2(x_1,\omega)]\approx  \frac{\rho}{\Pe^2}(x_1-\tfrac{1}{2} x_1^2) \mathrm{e}^{-(\Da/\Pe)x_1}.
\end{equation}
In 2D, downstream influence can again be captured approximately by using the far-field approximation (\ref{eq:g2farfield}) of $\mathcal{G}_2$ in the first integrals of (\ref{VarC1_Uniform_delta}) and (\ref{ExpC2_Uniform_simplified}) (taking $\Pe \gg \max(1,\sqrt{\Da})$, $\varsigma\ll 1/\Pe$ and $M\rightarrow \infty$) to give
\begin{equation}
\label{eq:2dvarapprox}
\mathrm{Var}[\widehat{C}_1(x_1,\omega)]\approx 
    {\rho^2}\sqrt{\frac{x_1}{8 \Pe^3 \pi}} \mathrm{e}^{-(2\Da/\Pe)x_1}, \quad \mathbb{E}[\widehat{C}_2(x_1,\omega)]\approx\rho^2   \frac{\log(1/(\Pe\varsigma))}{2\pi Pe} x_1 \mathrm{e}^{-(\Da/\Pe)x_1}.
\end{equation}
In 3D, the same approach using (\ref{eq:3dffg}) yields
\begin{equation}
\label{eq:3dvarapprox}
    \mathrm{Var}[\widehat{C}_1(x_1,\omega)]\approx \frac{\rho^3}{8\pi \Pe} \log(x_1 \lambda \Pe)\mathrm{e}^{-2(\Da/\Pe)x_1},\quad \mathbb{E}[\widehat{C}_2(x,\omega)]\approx  \frac{\rho^3}{4\pi^{3/2}\varsigma \Pe}x_1 \mathrm{e}^{-(\Da/\Pe)x_1},
\end{equation}
where $\lambda=O(1)$ is a constant that is not determined to this order and the variance expression is not valid near the inlet, when $x_1 \Pe=O(1)$. 

Integrals \chg{(\ref{eq:expc1}--\ref{eq:mc2})} were determined numerically using the solver given in \citet{Hosea2021integralN.m}, using 
adaptive quadrature functions in MATLAB. The domain $[0,1]\times[-3,3]$ was discretised with $251\times 1501$ points.  In 1D, approximations using the free-space Green's function were reduced to forms shown in Appendix~\ref{app:green}.  The \chg{asymptotic} region of influence of the 2D Green's functions (Figure~\ref{fig:GreensFunctionlengthscales}d) was used to identify sufficient domains of integration to ensure convergence.  

\subsection{Defining the effective Damk{\"o}hler number}
\label{sec:daeff}

In addition to calculating the mean correction directly via (\ref{ExpC2_Uniform}), we consider how the homogenization problem can be adjusted to capture the leading-order effect of disorder.  We seek the constant $\Da_{\eff}$ such that the solution of
\begin{equation} \label{GovEqnsEff}
\nabla^2_{3D} C  - \Pe C_{x_1} - \Da_{\eff} C = 0, \quad
\mathcal{B}_3C = \{1, 0, 0, 0, 0, 0\}
\end{equation}
approximates $\mathbb{E}[C(\mathbf{x};\omega)]$ to a suitable degree of accuracy. The exact solution of (\ref{GovEqnsEff}) is identical to the \chg{leading-order} homogenized solution given in (\ref{C_H}) but with $\Da$ replaced with $\Da_{\eff}$, namely
\begin{equation}\label{CH_eff}
C_H^{\chg{UR}}(\mathbf{x}) = C_H^{\chg{UR}}(x_1) = \dfrac{\Pe}{\Psi(1)} \Big( (2\Phi - \Pe)e^{\Phi(x_1-1)} + (2\Phi + \Pe) e^{\Phi(1-x_1)} \Big)e^{(\Pe/2)x_1},
\end{equation}
where $\Phi \equiv \sqrt{\Pe^2/4 + \Da_{\eff}}$ and $\Psi(x_1) \equiv (2\Pe\Phi + \Pe^2 + 2\Da_{\eff})e^{\Phi x_1} + (2\Pe\Phi - \Pe^2 - 2\Da_{\eff})e^{-\Phi x_1}$.  Writing $C(\mathbf{x};\omega) = C_H(x_1) + \widehat{C}(\mathbf{x};\omega)$, (\ref{GovEqnsEff}) can be rearranged to give  $\mathcal{L}_3\widehat{C}(\mathbf{x};\omega) = (\Da_{\eff}-\Da) \times (C_H(x_1) + \widehat{C}(\mathbf{x};\omega))$.
Assuming the correction $\widehat{C}(\mathbf{x}; \omega)$ is small compared to $C_H$, the linear operator can be inverted to give
\begin{equation}\label{correction}
\widehat{C}(\mathbf{x}) = (\Da_{\eff} - \Da) \int_{\mathcal{D}_3} \mathcal{G}_3(\mathbf{x} - \mathbf{x}') C_H(x_1') \, \mathrm{d}\mathbf{x}' + \dots,
\end{equation}
where the $\omega$ notation is dropped as the leading-order correction is deterministic.  We then rewrite (\ref{ExpC2_Uniform}) as
\begin{equation}\label{Exp1}
\mathbb{E}\left[\widehat{C}(\mathbf{x};\omega)\right] = \Da^2 \int_{\mathcal{D}_3} \int_{\mathcal{D}_3} \mathcal{G}_3(\mathbf{x} - \mathbf{x}') \mathcal{G}_3(\mathbf{x}' - \mathbf{x}'') \mathcal{K}_{\hat{g}}(\mathbf{x}', \mathbf{x}'') C_H(x_1'') \, \mathrm{d}\mathbf{x}' \, \mathrm{d}\mathbf{x}'' + \dots.
\end{equation}
Comparing this with (\ref{correction}) gives the approximate relation
\begin{gather}\label{Da_eff_gov_eqn}
\begin{split}
(\Da_{\eff} - \Da)\int_{\mathbb{R}^3} &\mathcal{G}_3(\mathbf{x} - \mathbf{x}') C_H(x_1') \, \mathrm{d}\mathbf{x}' \approx \Da^2 \int_{\mathbb{R}^3} \int_{\mathbb{R}^3} \mathcal{G}_3(\mathbf{x} - \mathbf{x}') \mathcal{G}_3 \widehat{\mathcal{K}}_{\hat{g}}(\mathbf{x}' - \mathbf{x}'') C_H(x_1'') \, \mathrm{d}\mathbf{x}' \, \mathrm{d}\mathbf{x}'',
\end{split}
\end{gather}
where $\mathcal{G}_3\widehat{\mathcal{K}}_{\hat{g}}(\mathbf{x}' - \mathbf{x}'') \equiv \mathcal{G}_3(\mathbf{x}' - \mathbf{x}'')\widehat{\mathcal{K}}_{\hat{g}}(\mathbf{x}' - \mathbf{x}'')$. In (\ref{Da_eff_gov_eqn}), we have expanded the domain $\mathcal{D}_3$ to $\mathbb{R}^3$, a reasonable assumption when sufficiently far from boundaries and the decay lengthscale of $\mathcal{G}_3 \widehat{\mathcal{K}}_{\hat{g}}$ is sufficiently short. 

\chg{Exploiting the fact} that $\mathcal{K}_{\hat{g}}(\mathbf{x},\mathbf{y})=\widehat{\mathcal{K}}_{\hat{g}} (\mathbf{x}-\mathbf{y})$ depends on $\mathbf{x}'-\mathbf{x}''$ rather than $\mathbf{x}'$ and $\mathbf{x}''$ \chg{independently}, we can rewrite (\ref{Da_eff_gov_eqn}) as
\begin{equation}\label{convolution}
(\Da_{\eff} - \Da) \mathcal{G}_3 \ast C_H \approx \Da^2 \mathcal{G}_3 \ast (\mathcal{G}_3 \widehat{\mathcal{K}}_{\hat{g}}) \ast C_H,
\end{equation}
where $\ast$ denotes convolution. If the decay lengthscale in $\widehat{\mathcal{K}}_{\hat{g}}$ is sufficiently short, then $\mathcal{G}_3\widehat{\mathcal{K}}_{\hat{g}}$ resembles a $\delta$-function with the appropriate weight and is given by \citep{Noetinger2018Effective}
\begin{equation}\label{relation}
\mathcal{G}_3\widehat{\mathcal{K}}_{\hat{g}}(\mathbf{y}) \approx \delta(\mathbf{y}) \int_{\mathbb{R}^3} \mathcal{G}_3\widehat{\mathcal{K}}_{\hat{g}}(\mathbf{x}) \, \mathrm{d}\mathbf{x}.
\end{equation}
Fourier transforming (\ref{convolution}), dividing by the non-zero Fourier transform of $C_H$ and applying the inverse transform, we obtain 
\begin{equation} \label{Da_eff}
\Da_{\eff} \approx \Da\left(1 + \Da\int_{\mathbb{R}^3} \mathcal{G}_3\widehat{\mathcal{K}}_{\hat{g}}(\mathbf{x})\, \mathrm{d}\mathbf{x}\right).
\end{equation}
As the Green's function and covariance function are always negative and positive respectively, $\Da_{\mathrm{eff}}$ is smaller than $\Da$, implying that disorder in the sink distributions reduces solute uptake.

For a sink covariance function of the form $\sigma^2 \exp (-\vert\mathbf{x}-\mathbf{y}\vert^2/\ell^2)$, taking $\ell \rightarrow 0$ and accounting for the singularity in $\mathcal{G}_2$ and $\mathcal{G}_3$, we evaluate (\ref{Da_eff}) using methods given in Appendix~\ref{app:W} to give
\begin{equation}
\label{Effective_Uptake_G_2D} 
\Da_{\eff} \approx
\begin{cases}
\Da \left( 1-\sqrt{\pi} \Da \sigma^2 \ell/(2\phi) \right) & \mathrm{(1D)} \\
 \Da\left(1 - \tfrac{1}{4} \Da \sigma^2 \ell^2 \left( \gamma -  2\log \left( \phi \ell \right) \right) \right) & \mathrm{(2D)}  \\
 \Da\left(1 - \tfrac{1}{2} \Da \sigma^2 \ell^2 \right)
 & \mathrm{(3D)}
\end{cases}
\end{equation}
where we have included the corresponding 1D approximation using (\ref{Da_eff}).  
Recall that $\phi = \sqrt{\Pe^2/4 + \Da}$.  The correction to $\Da$ in (\ref{Effective_Uptake_G_2D}) is proportional to $\ell$ (1D), $\ell^2 \log \ell$ (2D) and $\ell^2$ (3D), showing how the difference between $\Da$ and $\Da_{\eff}$ decreases with dimension for fixed variance and fixed correlation length. In 1D and 2D the correction is proportional to $1/\phi$ and $\log(\phi)$ respectively, whereas in 3D $\phi$ does not appear in the correction, demonstrating how the impact of advection on the effective uptake decreases as the dimension size increases. 

For uniformly random sinks in 2D and 3D letting $L_s \rightarrow \infty$, we can now use (\ref{eq:5a}), noting that the variance depends on sink size, to obtain  
\begin{equation}\label{Effective_Uptake_UR_2D} 
\Da_{\eff} \approx 
\begin{cases}
\Da\left(1 - \dfrac{\rho^2 \Da}{4\pi} \left( \gamma -  2\log \left( 2 \phi \varsigma \right) \right) \right) & \mathrm{(2D)}\\
\Da\left(1 - \dfrac{\rho^3 \Da}{4\pi^{3/2} \varsigma} \right) & \mathrm{(3D)}.
\end{cases}
\end{equation}
Used in combination with (\ref{CH_eff}), $C_H^{\chg{UR}}$ offers a direct estimate for the mean concentration field $\mathbb{E}[C]$ for uniformly-random sink locations in 2D and 3D, as we illustrate below.

\section{Results}
\label{sec:3}

\begin{figure}[t!]
\centering
\subfloat[]{\includegraphics[width=.45\linewidth]{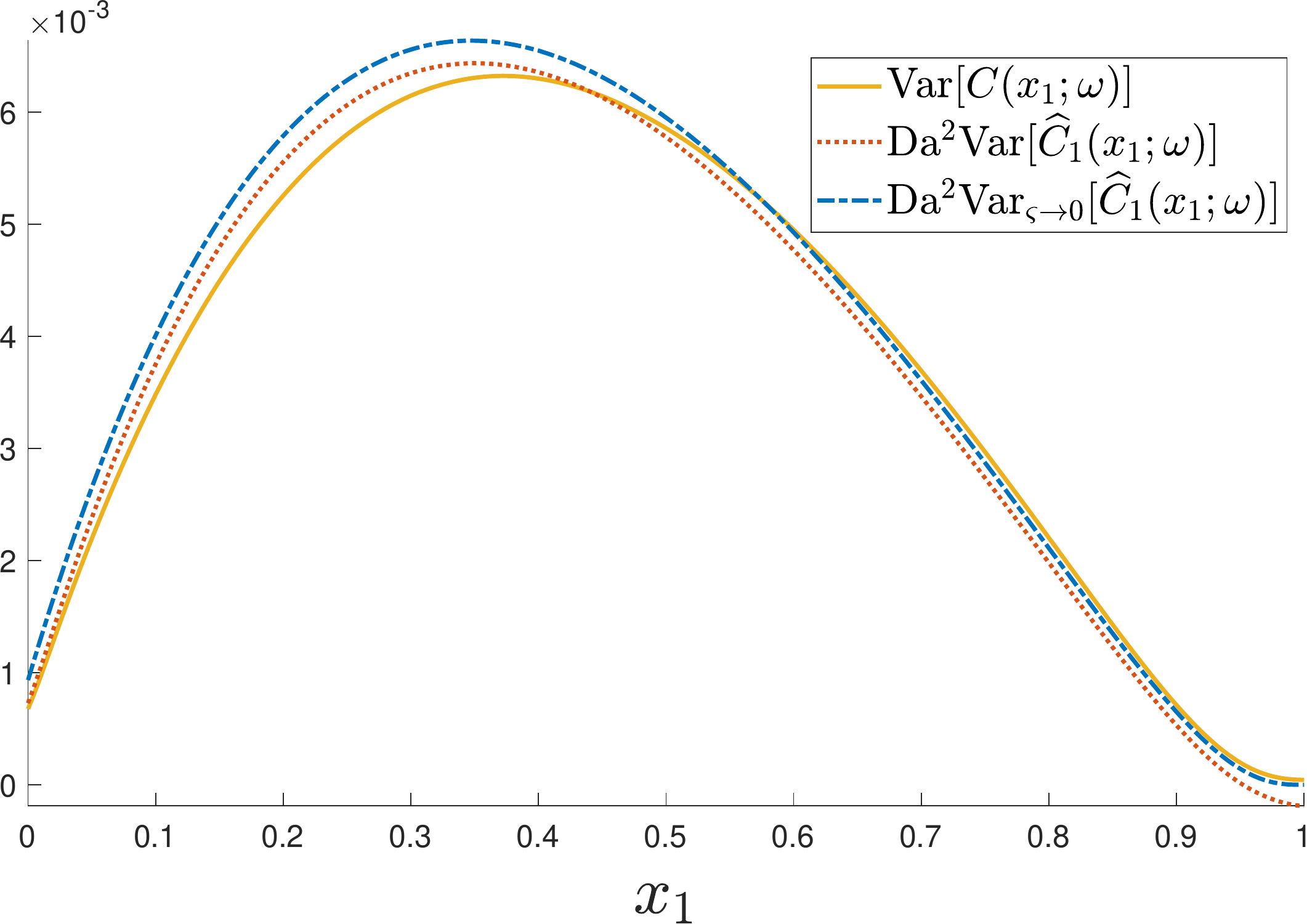}}
\hspace{0.5cm}
\subfloat[]{\includegraphics[width=.45\linewidth]{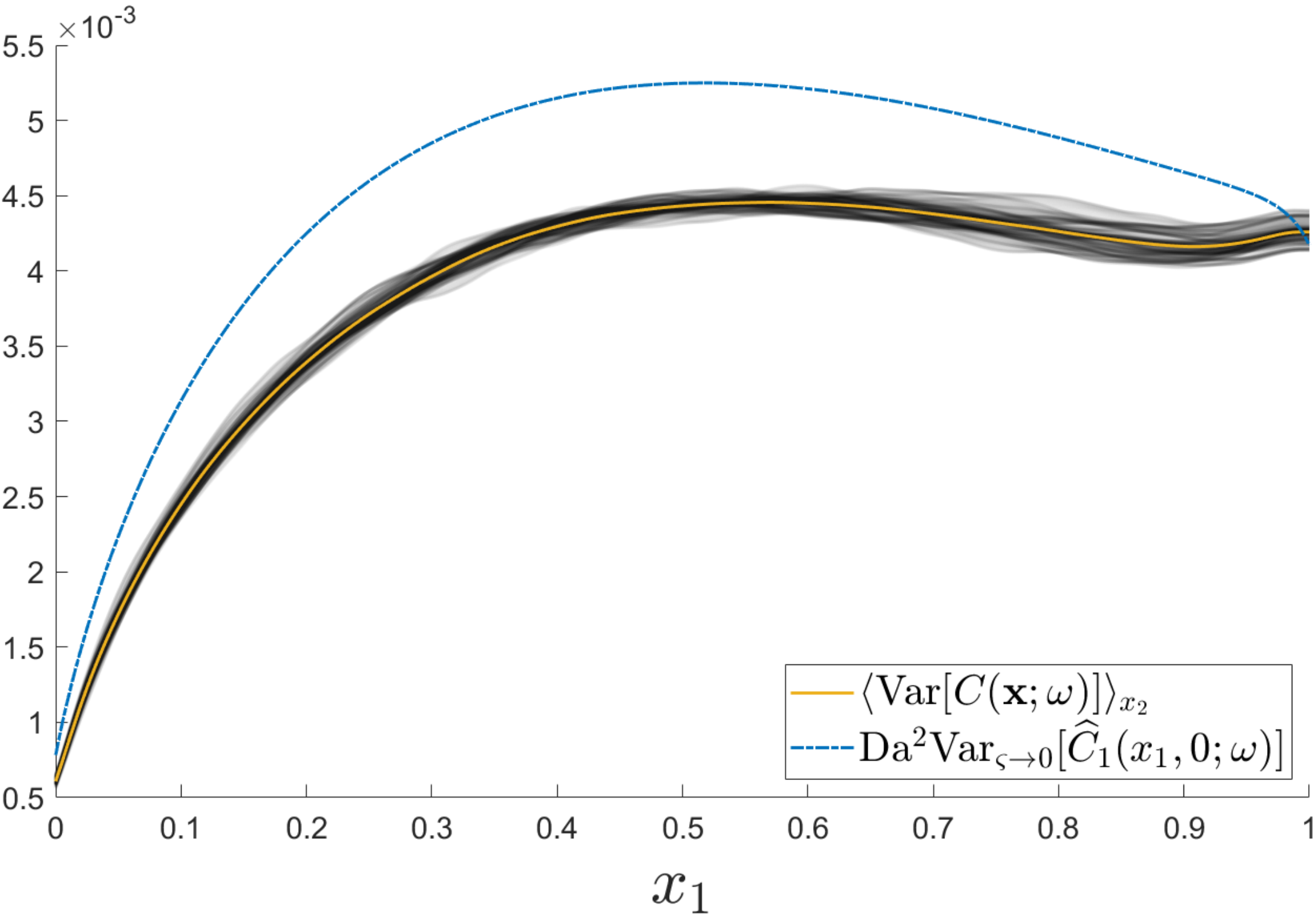}}
\caption{Variance of the concentration for $\rho=0.2$, $(\Pe, \Da) = (20, 10)$.  (a) 1D: $\Var[C(x_1;\omega)]$ (solid) represents the sample variance from $10^6$ Monte Carlo realisations, $\Var[\widehat{C}(x_1;\omega)]$ (dotted) is calculated using (\ref{CovC1_Uniform}) with $\varsigma=0.01$; $\Var_{\varsigma \rightarrow 0}[\widehat{C}(x_1;\omega)]$ (\chg{dot-}dashed) is calculated using (\ref{VarC1_Uniform_delta}). (b) 2D:
The cloud plot (grey) shows the sample variance for $x_2 =-2,-1.996, \dots, 2$ from Figure \ref{fig:Realisations_2D_L3}(c), the average of these variances over $x_2$ [$\langle \Var[C(\mathbf{x};\omega)] \rangle_{x_2}$, \chg{solid}] and the $\delta$-function approximation of the variance from (\ref{VarC1_Uniform_delta}) [$\Var_{\varsigma \rightarrow 0}[\widehat{C}_1(x_1,0;\omega)]$, \chg{dot-}dashed].  Sample variances are calculated from $10^4$ Monte Carlo realisations. }
\label{fig:OneDimensionVariance}
\end{figure}

The variance of the concentration field in 1D and 2D is illustrated in Figure~\ref{fig:OneDimensionVariance}.  The variance is smooth in both cases, due to strong mixing of sink locations over realisations.  In 1D, because exactly $N$ sinks are encountered along the domain, the concentration at the outlet is strongly constrained (as it was in Figure~\ref{fig:RealisationsZeroDiffusion}), and the variance falls close to zero at the outlet. In 2D this constraint is weaker ($N$ sinks are encountered \textit{on average} between $x_1=0$ and $x_1=1$), so that the variance remains large at the outlet; (\ref{eq:2dvarapprox}a), for example, predicts that the 2D variance is largest at the outlet for $4\Da<\Pe$. The cloud plot in Figure~\ref{fig:OneDimensionVariance}(b) demonstrates the magnitude of the sampling error from $10^4$ 2D simulations, and the independence of the transverse coordinate $x_2$ (Figure \ref{fig:Realisations_2D_L3}c). 

Figure~\ref{fig:OneDimensionVariance}(a) shows how the variance in 1D predicted by (\ref{eq:moments}b) matches closely with the sample variance taken from Monte Carlo simulations.  In 1D, the limit $\varsigma\rightarrow 0$ can be taken straightforwardly, using (\ref{VarC1_Uniform_delta}), and it provides a good approximation to the sample variance and the full integral (\ref{CovC1_Uniform}), while overpredicting the predicted variance uniformly.  The approximation (\ref{eq:1dvarapprox}a), using the leading-order approximation of the free-space Green's function for $\Pe\gg 1$, captures the shape of the variance well but over-predicts its maximum (predicting 0.0081 at $x_1\approx 0.38$ for the chosen parameter values, capturing its $x_1$-location well but over-estimating its value 0.0063 by almost 30\%).  In 2D, numerical evaluation of (\ref{CovC1_Uniform}) is expensive so we show only the simplified approximation (\ref{VarC1_Uniform_delta}), which overestimates the sample variance by approximately 10\% (due to neglect of finite sink size) but captures the overall features reasonably well.   The cruder prediction (\ref{eq:2dvarapprox}a) is also effective: it predicts the maximum variance at $x_1=\Pe/(4\Da)$ (for $\Pe<4\Da$) with value $\rho^2 \Da^2/\sqrt{16 \Pe^3 \pi e}$; the prediction $(0.5, 0.0038)$ underestimates the sample variance 0.0045 by about 15\%.  

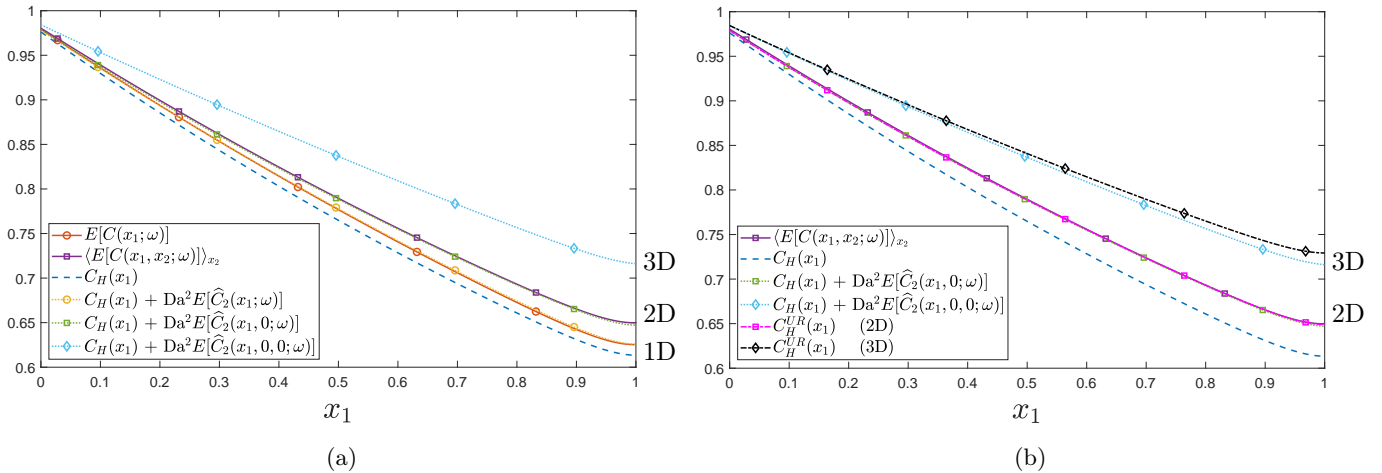
\begin{figure}[t!]
	\centering
	\subfloat[]
	{\begin{tikzpicture}
		\draw (0,0) node[above   right]
	{\includegraphics[width=.45\linewidth]{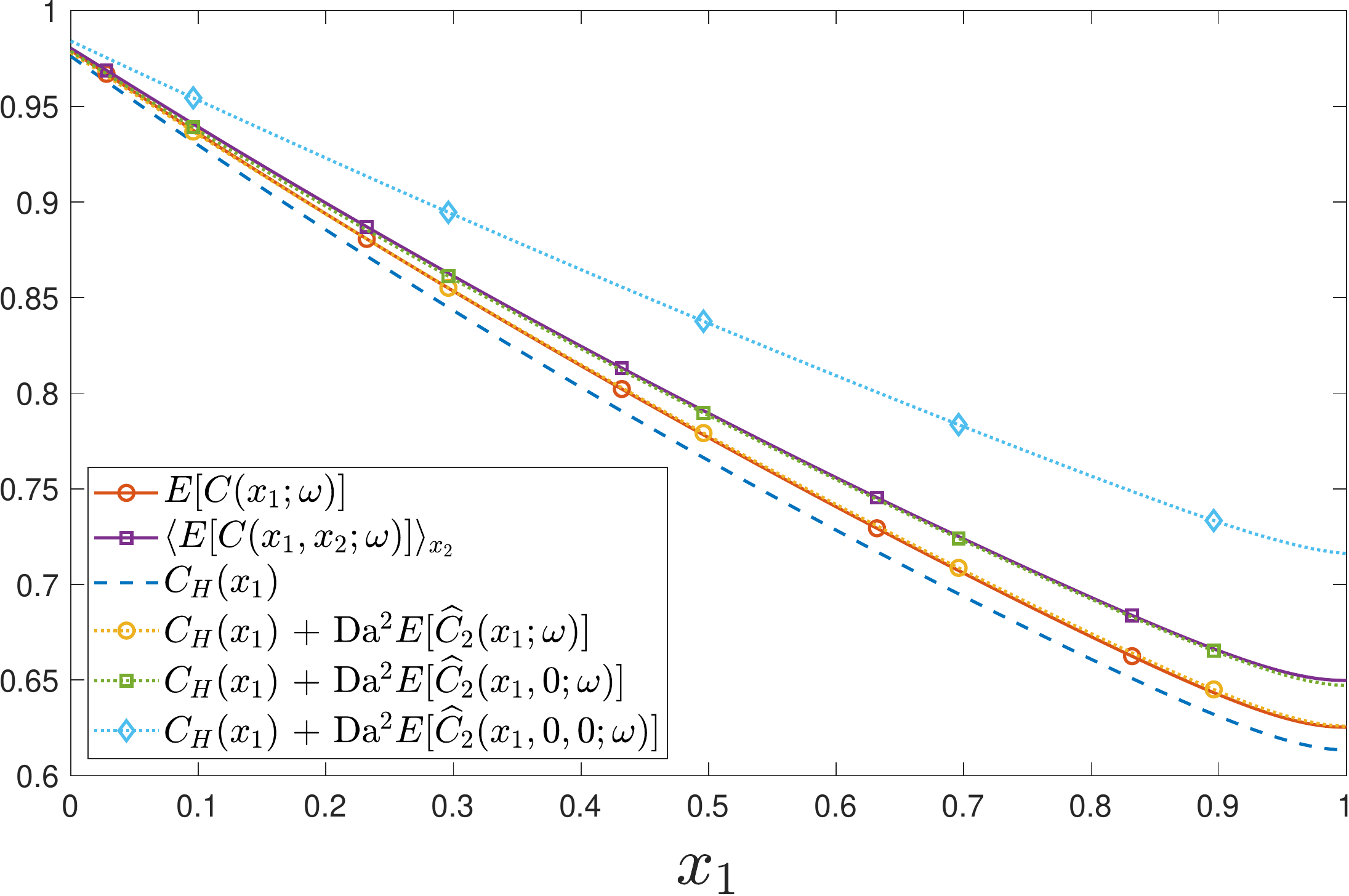}};
		\node[xshift=8.8cm, yshift=1.1cm]{\small{1D}};
		\node[xshift=8.8cm, yshift=1.6cm]{\small{2D}};
		\node[xshift=8.8cm, yshift=2.3cm]{\small{3D}};
		\end{tikzpicture}}
	\subfloat[]{\begin{tikzpicture}
		\draw (0,0) node[above   right]
	{\includegraphics[width=.45\linewidth]{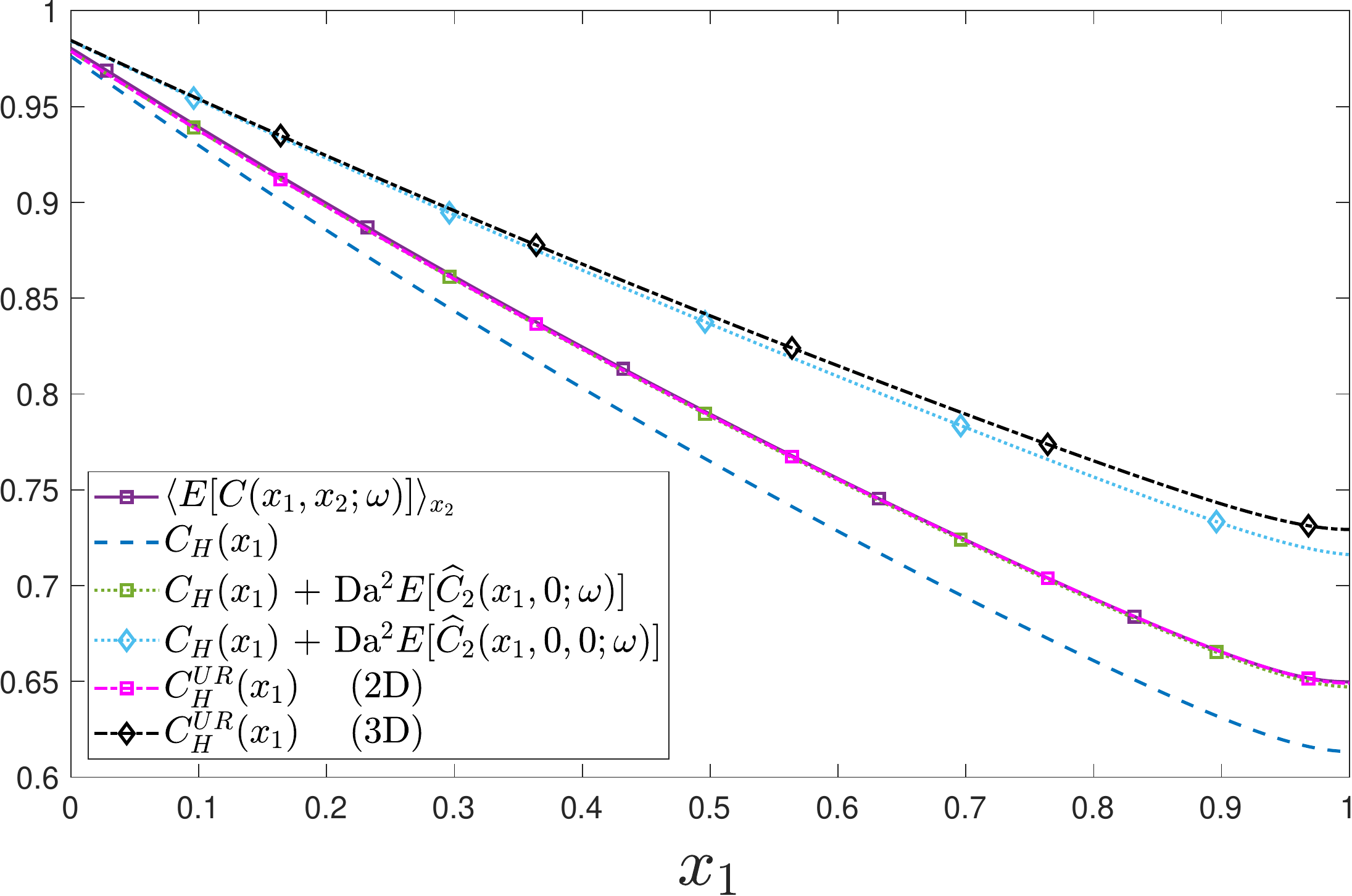}};
		\node[xshift=8.8cm, yshift=1.6cm]{\small{2D}};
		\node[xshift=8.8cm, yshift=2.3cm]{\small{3D}};
		\end{tikzpicture}} 
	\caption{Expected concentrations. Circles, squares and diamonds represent 1D, 2D and 3D domains respectively. (a) dashed and \chg{dotted} lines represent the \chg{leading-order} homogenized solution $C_H(x_1)$ \chg{plus} the approximation $\mathrm{Da}^2\mathbb{E}[\widehat{C}_2(\bm{x};\omega)]$ using (\ref{ExpC2_Uniform_simplified}). Solid lines represent the sample expectation, using $10^6$ realisations (1D) [$\mathbb{E}[\widehat{C}_2(x_1;\omega)]$] and $10^4$ realisations (2D) with $\mathcal{D}_2^s = [0, 1] \times [-2.5,2.5]$, averaging over $x_2 = -2, -1.996, \dots, 2$ [$\langle \mathbb{E}[\widehat{C}_2(x_1,x_2;\omega)]\rangle_{x_2}$], for $\rho = 0.2$, $\varsigma=0.01$ and $(\Pe, \Da) = (20, 10)$.
	(b) As in (a), with 2D and 3D effective uptake approximations in magenta and black respectively.  $C_H^{{UR}}$ was calculated using  (\ref{Effective_Uptake_UR_2D}) and (\ref{CH_eff}).}
	\label{fig:ExpectationUniform}
\end{figure}

Predictions of the ensemble mean concentration field are illustrated in Figure~\ref{fig:ExpectationUniform}(a). $\mathbb{E}[ \widehat{C}_2(\mathbf{x},\omega)]$ is  a smooth function of $x_1$, given by (\ref{ExpC2_Uniform_simplified}), and agrees well with the sample mean in 1D and 2D (stochastic simulations in 3D were not undertaken).  The correction compensates for the \chg{leading-order} homogenized solution over-predicting uptake.  The corrections grow with dimension, particularly through the factors $\beta_n$ from (\ref{beta_n}) as $\varsigma \rightarrow 0$.  In 2D and 3D when taking the limit $M \gg \rho$ (i.e. $L_s$ is asymptotically large), $\mathbb{E}[\widehat{C}_2(\mathbf{x};\omega)]$ can be simplified as the second integral becomes asymptotically small. Therefore the computational expense of calculating the correction is further reduced to solving one simple 1D integral.  The simpler estimate (\ref{eq:1dvarapprox}b) places the maximum 1D correction within the domain (but downstream of the maximum variance), of $O(\rho \Da^2/\Pe^2)$.  The 2D and 3D estimates (\ref{eq:2dvarapprox}b,c) place the maximum correction within the domain for $\Da>\Pe$, but at the outlet otherwise (as in Figure~\ref{fig:ExpectationUniform}), although they do not capture the weak boundary layer near $x_1=1$ evident in the figure.  

The mean concentration in 2D and 3D predicted using the $\Da_{\mathrm{eff}}$ approximation (\ref{Effective_Uptake_UR_2D}) is shown in Figure~\ref{fig:ExpectationUniform}(b).  The correction to $C_H$ grows with dimension, as expected, due to the increasingly large concentration fluctuations near each (regularised) sink.  The approximation provides close agreement to Monte Carlo sampling in 2D, and to the prediction (\ref{ExpC2_Uniform_simplified}) in 2D and 3D.  (Monte Carlo simulations in 3D were not attempted.)


\section{Discussion}
\label{sec:4}

This study has characterised the impact of spatial disorder on a transport process described by a linear advection/diffusion/uptake equation, assuming a uniformly random distribution of isolated sinks.  Using a \chg{leading-order} homogenization approximation (\ref{C_H_eqn}, \ref{C_H}) as a starting point, corrections were computed that describe the likely size of solute fluctuations around a mean field in a particular realisation, and the correction due to disorder when evaluating the ensemble mean concentration.  Bearing in mind the limitations of using statistical moments to characterise non-Gaussian concentration fields (Figure~\ref{fig:RealisationsZeroDiffusion}), we used a moments-based expansion to relate the mean and covariance of the sink distribution to the mean and covariance of the solute field (\ref{eq:moments}).  The first two moments of the sink distribution, when sinks are distributed uniformly randomly (\ref{Cov_g_UR_n_extra}), show an important distinction between 1D and higher dimensions, namely that in a sufficiently wide domain in 2D and 3D the correlation length of the sink covariance is set by the sink width (\ref{eq:5a}). Simulations (Figure~\ref{fig:Realisations_2D_L3}) reveal the multiscale nature of the problem: despite large concentration fluctuations in the neighbourhood of individual sinks in an individual realisation, ensemble averaging leads to smooth moments of the solute distribution with primary dependence only on a single spatial coordinate. Nevertheless, moments demand calculation of high-dimensional integrals, which we simplified by replacing the exact Green's function with its (explicit) free-space form, confining quadrature to appropriate regions of influence (Figure~\ref{fig:GreensFunctionlengthscales}d), replacing the regularised sink distribution (where possible) with its $\delta$-function approximation, and integrating over lateral dimensions using identities such as (\ref{GF-relation}).  This allowed accurate predictions of concentration means (Figure~\ref{fig:ExpectationUniform}a) in 1D and 2D, and of variance in 1D (Figure~\ref{fig:OneDimensionVariance}); the over-prediction of solute variance in 2D would likely be corrected by use of the regularised sink distribution, albeit using more expensive quadrature.  Cruder analytical estimates (\ref{eq:1dvarapprox}, \ref{eq:2dvarapprox}, \ref{eq:3dvarapprox}) were achieved by neglecting any upstream influence of one sink on another. 

For vanishingly small sinks (the limit $\varsigma\rightarrow 0$), concentration fields are discontinuous in 1D (Figure~\ref{fig:RealisationsZeroDiffusion}), and have $\log(1/\varsigma)$ and $1/\varsigma$ singularities in 2D and 3D respectively.  These appear both in corrections to the ensemble-averaged mean concentration (\ref{eq:mc2}) and in the effective Damk\"ohler number (\ref{Effective_Uptake_UR_2D}) that can be used in a modified homogenization approximation in 2D and 3D.  The latter approximation cannot be applied for uniformly random sinks in 1D, because the sink locations are correlated over the whole domain; however it can be applied when sinks are described by a Gaussian process with sufficiently short correlation length (\ref{Effective_Uptake_G_2D}).  For sink distributions of fixed variance, the impact of disorder falls as the sink correlation length vanishes (\ref{Effective_Uptake_G_2D}); however for uniformly random sinks in 2D and 3D the variance of the equivalent Gaussian process rises as $\varsigma$ falls (\ref{eq:5a}), \chg{contributing to the reduction in uptake captured in (\ref{Effective_Uptake_UR_2D}). As reported by \cite{Russell2020Homogenization} and \cite{Price2021}, mean correctors derived assuming periodic sink distributions show different dependence on parameters to those reported in (\ref{eq:1dvarapprox}b--\ref{eq:3dvarapprox}b).  For example, considering the expansion (\ref{expansion}), the dominant corrector in the deterministic periodic problem appears at $O(\mathrm{Da})$ and shares the wavelength of the microstructure, whereas the dominant correction in the uniformly-random case is stochastic with smooth variance ((\ref{eq:1dvarapprox}a--\ref{eq:3dvarapprox}a), Figure~\ref{fig:OneDimensionVariance}) with the mean correction appearing at $O(\mathrm{Da}^2)$ (Figure~\ref{fig:ExpectationUniform}).}

This study has a number of obvious extensions, prominent among which is consideration of other types of spatial disorder.  For flow in porous media, one expects the flow field to have disorder that correlates appreciably with the disorder in the sink distribution \cite{Jin2016Statistics}.  The present perturbative approach provides a route for understanding the contributions of flow, sinks and their combination to solute distributions, \chg{and it will be interesting to evaluate the present approach against predictions of existing studies of reactive transport in porous media relying either on periodicity assumptions \cite{Allaire2007Homogenization, hornung1991, mauri1991} or averaging procedures \cite{quintard1993}}.  Other obvious factors to consider include unsteady effects, variability in sink strength (considered in 1D by \cite{RussellJensenGalla16}) and the impact of a non-linear uptake kinetics (considered by \citet{Dalwadi2020systematic} using two-scale homogenization).  As demonstrated by \citet{ChernDrydenJensen12} and others, the statistical properties of the underlying spatial disorder interact with the physical lengthscales associated with transport processes to give a range of possible outcomes.  The present study illustrates some of the challenges in stepping away from traditional two-scale approaches towards non-local calculations, drawing attention to the need for efficient schemes for high-dimensional quadrature in order to characterise uncertainty.




We can revisit the expansion (\ref{expansion}) and use evidence that terms $\Da \widehat{C}_1$ or $\Da^2 \widehat{C}_2$ become comparable in magnitude to $C_H$ as evidence of the breakdown of a homogenization approximation. In 1D,  based on the estimates in $(\ref{eq:1dvarapprox})$, the restriction $\Pe\gg \max(1,\sqrt{\Da})$ must be extended to $\Pe\gg\max(1,\sqrt{\Da}, \rho^{1/2}\Da)$, which holds along the distinguished limit $\Pe\sim\Da$ for arbitrarily large $\Pe$.  The parameter $\Da^2 \rho/\Pe$, measuring the magnitude (relative to $C_H$) of the fluctuation variance and the correction to the mean, takes the value 0.05 in  Figure~\ref{fig:RealisationsZeroDiffusion} (with $\Pe\rightarrow \infty$, but with $\Da \rho^{1/2}/\Pe=S_1/\rho^{1/2}$; see Appendix~\ref{app:A}) and Figures~\ref{fig:OneDimensionVariance}(a) and \ref{fig:ExpectationUniform}(a). In these examples, fluctuations with standard deviation of order 20\% dominate the correction to the mean, of order 5\%.  In 2D and 3D however, the range of validity of the approximation is reduced and the correction to the mean (that grows with diminishing sink size) overtakes the fluctuations as the dominant correction.  In 3D, we require $\Pe\gg\max(1,\sqrt{\Da},\Da^2\rho^3/\varsigma)$ (for $\rho^3\ll \varsigma \ll \rho \ll 1$), which confines the distinguished limit to $1\ll \Pe\sim\Da \ll \varsigma/\rho^3$.  The example shown in Figure \ref{fig:ExpectationUniform} has $\varsigma/\rho^3=1.25$: as the figure indicates, the predicted correction to the mean is sufficiently large to call into question the validity of the homogenization approximation in this case.  In 2D, the constraint on the distinguished limit is $1\ll \Pe\sim\Da \ll 1/(\rho^2 \log(\rho^2/\varsigma))$: the example in Figures~\ref{fig:Realisations_2D_L3}, \ref{fig:OneDimensionVariance}(b) and \ref{fig:ExpectationUniform}(a) with $\rho^{-2}=25$ therefore sits at this upper threshold, although the predicted corrections are still effective.


\appendix

\section{1D concentration profiles with zero diffusion}
\label{app:A}
\renewcommand{\theequation}{A.\arabic{equation}}

Let $\xi_i$ ($i=1,\dots, N$) denote point sink locations, distributed as order statistics $U_{j:N}$ taken from a uniform distribution $U \sim \mathcal{U}(0,1)$ with \chg{probability density function} (pdf) $\pi_{U}(x)=1$ for $0 \le x \le 1$ and zero otherwise.  Each sink location follows a Beta distribution \cite{Arnold1992first} such that $\xi_j \sim \beta(j, N-j+1)$, where $j=1,\dots,N$. Here $\beta(x,y) \equiv t^{x-1} (1-t)^{y-1} / \mathrm{B}(x,y)$, where B$(x,y) \equiv \Gamma(x)\Gamma(y)/\Gamma(x+y)$.  The cumulative distribution function (cdf) $F_{\xi_j}(x) = \mathbb{P}(\xi_j \le x)$ is given by the regularised incomplete beta function
\begin{equation} \label{sinklocationCDFuniform}
F_{\xi_j}(x) = I_{x}(j,N-j+1) = \dfrac{\int_{0}^{x}t^{j-1} (1-t)^{N-j}\mathrm{d}t}{\mathrm{B}(j,N-j+1)}.
\end{equation}

The 1D concentration distribution that falls by a factor $1/(1+S_1)$ at each sink from its inlet value $C_0=1$ (Figure~\ref{fig:RealisationsZeroDiffusion}) satisfies
\begin{equation}\label{ConcZeroDiffusion}
C(x) = C_0 - S_1 {\textstyle \sum_{j=1}^{N}} C_j H(x-\xi_j), \quad 
C_j \equiv (1+S_1)^{-j}C_0.
\end{equation}
(This problem can be defined as a limit of the 1D form of (\ref{GoverningEquation}), with $\Pe\rightarrow \infty$ taking $S_1=\Da \rho/\Pe$ with $\Da/\Pe=O(1)$.)  The probability of being at concentration $C_j$ for some given $x$ is $\mathbb{P}(C_j ; x) = \mathbb{P}(\xi_j < x < \xi_{j+1}) = F_{\xi_j}(x) - F_{\xi_{j+1}}(x)$ for $j=1,\dots, N-1$, with $\mathbb{P}(C_0 ; x) = \mathbb{P}(\xi_1 > x) = 1 - F_{\xi_1}(x)$, $\mathbb{P}(C_N ; x) = \mathbb{P}(\xi_N < x) = F_{\xi_N}(x)$.
Therefore the expectation $\mathbb{E}[C(x)] = C_0 \mathbb{P}(C_0 ; x) + \dots + C_N \mathbb{P}(C_N ; x)$ becomes
\begin{align}
\mathbb{E}[C(x)] &= C_0 (1 - F_{\xi_1}(x)) + \sum_{j=1}^{N-1} C_j (F_{\xi_j}(x) - F_{\xi_{j+1}}(x)) + C_N F_{\xi_N}(x),
 \nonumber \\ &
 = C_0 + \sum_{j=1}^{N} (C_j - C_{j-1}) F_{\xi_j}(x) = 1 - S_1 \sum_{j=1}^{N}  \frac{I_{x}(j,N-j+1)}{\left(1+S_1 \right)^{j}}. 
 \label{ExpZeroDiffusionUniform}
\end{align}
The variance $\Var[C(x)] = \sum_{i=0}^N C_i^2 \mathbb{P}(C_i;x) -(\sum_{i=0}^N C_i \mathbb{P}(C_i; x))^2$ satisfies
\begin{align}
\Var[C(x)]& = (C_0)^2 + \sum_{j=1}^{N} ((C_j)^2 - (C_{j-1})^2) F_{\xi_j}(x) - (C_0 + \sum_{j=1}^{N} (C_j - C_{j-1}) F_{\xi_j}(x))^2 \nonumber \\
\label{VarZeroDiffusionUniform}
 &= \sum_{j=1}^{N} \left( C_j + C_{j-1} - 2 C_0  - \sum_{i=1}^{N} (C_i - C_{i-1})F_{\xi_i}(x)\right)(C_j - C_{j-1})F_{\xi_j}(x) \\
 &= S_1\sum_{j=1}^{N}\Bigg( 2 - \frac{(2+S_1)}{(1+S_1)^{j}} - S_1\sum_{i=1}^{N} \frac{I_{x}(i,N-i+1)}{(1+S_1)^{i}}\Bigg) 
 \frac{I_{\varepsilon x}(j,N-j+1)}{(1+S_1)^{j}}.
\end{align}
$\mathbb{E}[C(x)]$ and $\mathrm{Var}[C(x)]$ are plotted in Figure~\ref{fig:RealisationsZeroDiffusion} using (\ref{sinklocationCDFuniform}).

The cdf of the concentration $C_j$ is given by
\begin{equation}\label{ZeroDiffusionCDF}
F_{C_j}(C) = \mathbb{P}(C_j \le C(x) ; x) = \mathbb{P}(\xi_j > x) = 1 - F_{\xi_j}(x) \quad (j=1,\dots, N).
\end{equation}
Let the cdf take a value $F_{C_j}(C) = r$. Then (\ref{ZeroDiffusionCDF}) can be inverted to give the corresponding sink locations as
\begin{equation}\label{ZeroDiffusionCDFx}
\breve{\xi}_j = F^{-1}_{\xi_j}(1 - r) = \varepsilon^{-1} I^{-1}_{r}(j,N-j+1) \quad
(j=1, \dots, N).
\end{equation}
We can therefore use (\ref{ConcZeroDiffusion}) to find the cdf credible intervals as
\begin{equation}\label{ZeroDiffusionCIUniform}
CI(x;r) = C_0 - S_1 C_0 \sum_{j=1}^{N}  \frac{H(x -  I^{-1}_{r}(j,N-j+1))}{(1+S_1)^{j}}.
\end{equation}
Credible intervals which ensure 95\% of concentration profiles are contained between the two bounds are shown in Figure~\ref{fig:RealisationsZeroDiffusion} using $r = 0.025$ and $r = 0.975$ in (\ref{ZeroDiffusionCIUniform}); the median is evaluated using $r=0.5$.
Credible intervals respect the  requirement that the concentration is bounded between $C_N$ at the outlet and $C_0$ at the inlet, demonstrating that the solute distribution is non-Gaussian.

\section{Moments of the sink distribution}
\label{sec:SinkFunctionMoments}
\renewcommand{\theequation}{B.\arabic{equation}}

Let sink locations in 3D be prescribed by a multivariate uniform distribution, with position vectors $\bm{\xi}_{\mathbf{i}_3} = (\xi_i,\xi_j,\xi_k)$ such that $\xi_i \sim \mathcal{U}[0,1]$ and $\xi_j, \xi_k \sim \mathcal{U}[-L_s,L_s]$ for $i = 1, \dots, N$ and $j,k = -M, \dots, M$. Each continuous uniformly-random variable $\bm{\xi}_{\mathbf{i}_3}$ is independently and identically distributed with a pdf given by
\begin{align}\label{UniformPDF}
\pi_{\bm{\xi}_{\mathbf{i}_3}}(\mathbf{x}_{\mathbf{i}_3}) = \begin{cases}
\dfrac{1}{(2L_s)^2} = \dfrac{1}{\rho^2(2M+1)^2} \quad &\text{ for } \mathbf{x}_{\mathbf{i}_3} \in \mathcal{D}_3^s, \\
0 \quad &\text{ otherwise.}
\end{cases}
\end{align}
Using the definition of the sink function given in (\ref{g-discrete}), the expectation of $\hat{g}(\mathbf{x};\omega)$ is given by
\begin{align}\label{Exp_g_UR}
\mathbb{E}\left[\hat{g}(\mathbf{x};\omega)\right] &= \int_{\mathcal{D}_3} \int_{\mathcal{D}_3} \dots  \left(\rho^3\sum_{\mathbf{i}_3} F^{(3)}_{\varsigma}(\mathbf{x} - \mathbf{x}_{\mathbf{i}_3}) - 1 \right) \pi_{\bm{\xi}_{\mathbf{1}},\bm{\xi}_{\mathbf{2}},\dots} (\mathbf{x}_{\mathbf{1}},\mathbf{x}_{\mathbf{2}},\dots) \, \mathrm{d}\mathbf{x}_{\mathbf{1}} \mathrm{d}\mathbf{x}_{\mathbf{2}} \dots \nonumber\\
&= \rho^3\sum_{\mathbf{i}_3} \int_{\mathcal{D}_3} F^{(3)}_{\varsigma}(\mathbf{x} - \mathbf{x}_{\mathbf{i}_3})  \pi_{\bm{\xi}_{\mathbf{i}_3}}(\mathbf{x}_{\mathbf{i}_3}) \, \mathrm{d}\mathbf{x}_{\mathbf{i}_3} - 1 = 0.
\end{align}

To calculate the covariance $\mathcal{K}_{\hat{g}}(\mathbf{x},\mathbf{y}) = \mathbb{E}[\hat{g}(\mathbf{x};\omega) \hat{g}(\mathbf{y};\omega)] $ we can again use (\ref{UniformPDF}) to obtain
\begin{gather*}
\begin{split}
\mathcal{K}_{\hat{g}}(\mathbf{x},\mathbf{y})& = \rho^{6}\mathop{\sum_{\mathbf{i}_3}\sum_{\mathbf{j}_3}}_{\mathbf{i}_3 \ne \mathbf{j}_3} \int_{\mathcal{D}_3} \int_{\mathcal{D}_3} F^{(3)}_{\varsigma}(\mathbf{x} - \mathbf{x}_{\mathbf{i}_3}) F^{(3)}_{\varsigma}(\mathbf{y} - \mathbf{x}_{\mathbf{j}_3}) \pi_{\bm{\xi}_{\mathbf{i}_3},\bm{\xi}_{\mathbf{j}_3}} (\mathbf{x}_{\mathbf{i}_3},\mathbf{x}_{\mathbf{j}_3}) \, \mathrm{d}\mathbf{x}_{\mathbf{i}_3} \, \mathrm{d}\mathbf{x}_{\mathbf{j}_3} \\
& + \rho^{6}\sum_{\mathbf{i}_3}\int_{\mathcal{D}_3} F^{(3)}_{\varsigma}(\mathbf{x} - \mathbf{x}_{\mathbf{i}_3}) F^{(3)}_{\varsigma}(\mathbf{y} - \mathbf{x}_{\mathbf{i}_3}) \pi_{\bm{\xi}_{\mathbf{i}_3}}(\mathbf{x}_{\mathbf{i}_3}) \, \mathrm{d}\mathbf{x}_{\mathbf{i}_3} \\
&- \rho^3\sum_{\mathbf{i}_3} \int_{\mathcal{D}_3} F^{(3)}_{\varsigma}(\mathbf{x} - \mathbf{x}_{\mathbf{i}_3})  \pi_{\bm{\xi}_{\mathbf{i}_3}}(\mathbf{x}_{\mathbf{i}_3}) \, \mathrm{d}\mathbf{x}_{\mathbf{i}_3}
-\rho^3\sum_{\mathbf{j}_3} \int_{\mathcal{D}_3} F^{(3)}_{\varsigma}(\mathbf{y} - \mathbf{x}_{\mathbf{j}_3})  \pi_{\bm{\xi}_{\mathbf{j}_3}}(\mathbf{x}_{\mathbf{j}_3}) \, \mathrm{d}\mathbf{x}_{\mathbf{j}_3} + 1
\end{split}
\end{gather*}
which gives
\begin{equation}\label{Cov_g_UR}
\mathcal{K}_{\hat{g}}(\mathbf{x},\mathbf{y}) = \rho^3 \mathzapf{F}^{(3)}_{\varsigma}(\mathbf{x},\mathbf{y}) - \dfrac{\rho}{(2M+1)^2} \quad \mathrm{where}\quad
\mathzapf{F}^{(3)}_{\varsigma}(\mathbf{x},\mathbf{y}) \equiv \int_{\mathcal{D}_3} F^{(3)}_{\varsigma}(\mathbf{x} - \hat{\mathbf{x}}) F^{(3)}_{\varsigma}(\mathbf{y} - \hat{\mathbf{x}}) \, \mathrm{d}\hat{\mathbf{x}}.
\end{equation}
The function $\mathzapf{F}^{(3)}_{\varsigma}$ measures the overlap of the two functions $F^{(3)}_{\varsigma}(\mathbf{x} - \hat{\mathbf{x}})$ and $F^{(3)}_{\varsigma}(\mathbf{y} - \hat{\mathbf{x}})$
and is zero when $\mathbf{x}$ is sufficiently far from $\mathbf{y}$. 
When $F^{(3)}_{\varsigma}$ has a Gaussian structure  (\ref{F_varsigma}), we find that  
\begin{equation*}
\mathzapf{F}^{(3)}_{\varsigma}(\mathbf{x},\mathbf{y}) = I_2(x_1,y_1;\varsigma,\varsigma)I_2(x_2,y_2;\varsigma,\varsigma)I_2(x_3,y_3;\varsigma,\varsigma) = F^{(3)}_{\sqrt{2}\varsigma}(\mathbf{x} - \mathbf{y}),
\end{equation*}
where 
\begin{equation}
I_{2}\left(x, y ; \sigma_{x}, \sigma_{y}\right) \equiv \frac{1}{2 \pi \sigma_{x} \sigma_{y}} \int_{-\infty}^{\infty} \exp \left(-\frac{1}{2 \sigma_{x}^{2}}(\hat{x}-x)^{2}-\frac{1}{2 \sigma_{y}^{2}}(\hat{x}-y)^{2}\right) \mathrm{d} \hat{x}.
\end{equation}
This in turn gives the covariance of $\hat{g}$ as in (\ref{Cov_g_UR_n_extra}) for $n=3$; we can extend these results using similar calculations for $n=1$ and 2 dimensions,
noting that the number of sinks $(2M+1)^2/\rho$ becomes $(2M+1)^{n-1}/\rho$. 

\section{Green's functions}
\label{app:green}
\renewcommand{\theequation}{C.\arabic{equation}}


The exact Green's function in 1D $G(x_1,x_1')$  satisfies $\mathcal{L} G = \delta(x_1-x_1')$, $\mathcal{B}_1G = \{ 0 , 0 \}$, where 
$\mathcal{L} = \left(\partial_{x_1}\right)^2 - \Pe\partial_{x_1} - \Da$ and $\mathcal{B}_1 = \{\left( 1 - (1/\Pe)\partial_{x_1} \right)(\cdot) |_{x_1=0} , \, \partial_{x_1}(\cdot) |_{x_1=1}\}$.  
We define $G^{-}(x_1,x_1')$ and $G^{+}(x_1,x_1')$ such that 
\begin{equation*}
G(x_1,x_1') = \begin{cases}
G^{-}(x_1,x_1') & \text{if } 0 \le x_1 \le x_1' \le 1\\
G^{+}(x_1,x_1') & \text{if } 0 \le x_1' \le x_1 \le 1
\end{cases}
\end{equation*}
where
\begin{gather}\label{G_1D}
\begin{split}
G^{\pm}(x_1,x_1') &= -\left(\dfrac{1}{4 \phi \psi(1)}\right) e^{\frac{\Pe}{2}(x_1 - x_1')}
\Bigg( (2\phi + \Pe)^2 e^{\phi (\pm(x_1' - x_1) + 1)} \\
&\quad + (2\phi - \Pe)^2 e^{- \phi(\pm(x_1' - x_1) + 1)} + 4 \Da \left(e^{\phi (x_1 + x_1' - 1)} + e^{-\phi (x_1 + x_1' - 1)}\right) \Bigg).
\end{split}	
\end{gather}
It is convenient to re-express this to allow numerical evaluation when $\Pe^2 \gg \Da$.  We expand exponential terms to obtain
\begin{gather*}
\phi \approx \dfrac{\Pe}{2} + \dfrac{\Da}{\Pe} - \dfrac{\Da^2}{\Pe^3}, \quad
\exp\left(\pm\phi x_1\right) \approx  \left(1 \mp \dfrac{\Da^2}{\Pe^3}x_1 \right)\exp\left(\pm\left(\dfrac{\Pe}{2} + \dfrac{\Da}{\Pe}\right)x_1\right), \\
\psi(x_1) \approx  2\Pe^2 \left(1 + \dfrac{\Da}{\Pe^2}\left( 2 - \dfrac{\Da}{\Pe}x_1 \right) \right)\exp\left(\left(\dfrac{\Pe}{2} + \dfrac{\Da}{\Pe}\right)x_1\right),
\end{gather*}
giving
\begin{gather}
\begin{split}
\widetilde{G}^{-}(x_1,x_1') \approx &-\dfrac{1}{\Pe}e^{\left(\Pe + \dfrac{\Da}{\Pe}\right) (x_1-x_1')} 
+ \dfrac{\Da}{\Pe^3}\Bigg(\left(2 + \dfrac{\Da}{\Pe}(x_1-x_1')\right) e^{\left(\Pe + \dfrac{\Da}{\Pe}\right) (x_1-x_1')} \\
&- e^{\Pe(x_1-1) + \dfrac{\Da}{\Pe}(x_1+x_1'-2) }
- e^{-\Pe x_1' - \dfrac{\Da}{\Pe}(x_1+x_1')} \Bigg), \\
\widetilde{G}^{+}(x_1,x_1') \approx &-\dfrac{1}{\Pe}e^{\dfrac{\Da}{\Pe} (x_1'-x_1)} 
+ \dfrac{\Da}{\Pe^3}\Bigg(\left(2 + \dfrac{\Da}{\Pe}(x_1'-x_1)\right) e^{\dfrac{\Da}{\Pe} (x_1'-x_1)} \\
&- e^{\Pe(x_1-1) + \dfrac{\Da}{\Pe}(x_1+x_1'-2)} 
- e^{-\Pe x_1' - \dfrac{\Da}{\Pe}(x_1+x_1')} \Bigg). 
\end{split}
\end{gather}
From the first term in $G^{-}$ ($G^{+}$) we see a boundary layer of width approximately $1/\Pe$ ($\Da/\Pe$) exists upstream (downstream) of $x_1=x_1'$ (Figure~\ref{fig:GreensFunctionlengthscales}a). The final two terms in $G^{-}$ and $G^{+}$ account for the boundary conditions, which gives a boundary layer of width approximately $1/\Pe$ at the $x_1$-outlet and $x_1'$-inlet.


The $n$-dimensional free-space Green's function $\mathcal{G}_n (\mathbf{x}-\mathbf{x}')$ associated with (\ref{GoverningEquation}) satisfies $\mathcal{L}_nG^n=\delta(\mathbf{x}-\mathbf{x}')$ and decays in the far field.  Seeking a solution of the form $\mathcal{G}_n(\mathbf{x}) = e^{\lambda x_1} f(r)$ where $r=|\mathbf{x}|$ and setting $\lambda = \Pe/2$ leads to
\begin{equation} \label{Free-Space-GF}
\mathcal{G}_n(\mathbf{x}-\mathbf{x}') = - (2\pi)^{-n/2} \left( \dfrac{\phi}{|\mathbf{x} - \mathbf{x}'|} \right)^{n/2 - 1}K_{n/2 - 1}(\phi |\mathbf{x} - \mathbf{x}'|) \exp \left(\dfrac{\Pe}{2}(x_1-x_1')\right),
\end{equation}
where $\phi \equiv \sqrt{\Pe^{\,2}/4 + \Da}$ 
and $K_{\nu}$ represents the modified Bessel function of the second kind \citep{Tikhonov2013Equations}.
The free-space Green's function in 1D is readily evaluated, noting that  
$K_{\pm 1/2}(z) = \sqrt{{\pi}/{(2 z)}}\exp (- z)$, as
\begin{equation}\label{Free-Space-GF-1D}
\mathcal{G}_1(x_1-x_1') = - \dfrac{1}{2\phi}  \exp \left(\dfrac{\Pe}{2}(x_1-x_1') - \phi |x_1-x_1'| \right).
\end{equation} 
As illustrated in Figure \ref{fig:GreensFunctionlengthscales}(a,b), for $\Pe \gg \max(1,\sqrt{\Da})$, $\mathcal{G}_1$ decays on a short lengthscale $1/\Pe$ upstream of $x_1=x_1'$, and on a long lengthscale $\Pe/\Da$ downstream, but fails to capture additional boundary layers of width $1/\Pe$ in $G^+$ at the edges of the domain.

From (\ref{Free-Space-GF}), the 2D free-space Green's function is
\begin{equation}\label{Free-Space-GF-2D}
\mathcal{G}_2(\mathbf{x}-\mathbf{x}') = - \dfrac{1}{2\pi} K_0(\phi |\mathbf{x} - \mathbf{x}'|) \exp \left(\dfrac{\Pe}{2}(x_1-x_1')\right).
\end{equation}
Thus 
\begin{equation} \label{Free-Space-GF-2D_near_field}
\mathcal{G}_2(\mathbf{x}-\mathbf{x}') \approx 
\begin{cases}
\dfrac{1}{2 \pi} \log(\phi |\mathbf{x} - \mathbf{x}'|) & 
\phi\vert\mathbf{x}-\mathbf{x}'\vert\ll 1, \\
\mathcal{G}_2(\mathbf{x}-\mathbf{x}') \approx - \dfrac{1}{2} \sqrt{\dfrac{1}{2 \pi \phi |\mathbf{x} - \mathbf{x}'|}} \exp \left(\dfrac{\Pe}{2}(x_1-x_1') - \phi |\mathbf{x} - \mathbf{x}'|\right) &
\phi\vert\mathbf{x}-\mathbf{x}'\vert\gg 1.
\end{cases}
\end{equation}
Along $x_2=x_2'$, when $\Pe\gg \max(1,\Da)$, $\mathcal{G}_2$ decays over the same lengthscales as $\mathcal{G}_1$.  Along $x_1=x_1'$, $\mathcal{G}_2$ decays over a distance $1/\Pe$ in the $x_2$ direction.  
The asymptotic shape of the wake in the far field is revealed by rescaling using $x_1-x_1' = (\Pe/\Da)X_1$ and $x_2-x_2' = (1/\sqrt{\Da})X_2$ for $X_1, X_2 = \mathcal{O}(1)$. We can then approximate
(\ref{Free-Space-GF-2D_near_field}b) as
\begin{equation}
\label{eq:g2farfield}
\mathcal{G}_2(\mathbf{x}-\mathbf{x}') \approx - \dfrac{1}{2\Pe} \sqrt{\dfrac{\Da}{\pi X_1}} \exp 
\left( - X_1-\frac{X_2^2}{4X_1}+\dots \right)
\end{equation}
for $\Pe\gg \sqrt{\Da}$.  The argument of the exponential identifies the approximately elliptical shape of concentration contours, 
as sketched in Figure~\ref{fig:GreensFunctionlengthscales}(c).


The 3D free-space Green's function is
\begin{equation}\label{Free-Space-GF-3D}
\mathcal{G}_3(\mathbf{x}-\mathbf{x}') = - \dfrac{1}{4 \pi |\mathbf{x} - \mathbf{x}'|}  \exp \left(\dfrac{\Pe}{2}(x_1-x_1') - \phi |\mathbf{x}-\mathbf{x}'| \right).
\end{equation}
This has near-field form
\begin{equation}\label{Free-Space-GF-3D_near_field}
\mathcal{G}_3(\mathbf{x} - \mathbf{x}') \approx - \dfrac{1}{4 \pi r},\quad \mathrm{as}~r=\vert\mathbf{x}-\mathbf{x}'\vert\rightarrow 0
\end{equation}
while the far-field structure for $\Pe\gg \max(1,\sqrt{\Da})$ can be written
\begin{equation}
\label{eq:3dffg}
    \mathcal{G}_3(\mathbf{x}-\mathbf{x}')\approx - \frac{1}{4\pi (x_1-x_1')}\exp\left[-\frac{\Da}{\Pe}(x_1-x_1')-\Pe\frac{(x_2-x_2')^2+(x_3-x_3')^2)}{4(x_1-x_1')} \right],
\end{equation}
with lengthscales resembling those illustrated in Figure~\ref{fig:GreensFunctionlengthscales}(c).

\section{Evaluating integrals}
\label{app:Y}
\renewcommand{\theequation}{D.\arabic{equation}}

In 1D, (\ref{Free-Space-GF-1D}) with $x_1''=x_1' + \varsigma u$ gives
$\mathcal{G}_1(x_1'-x_1'') = -({1}/{(2\phi)})  \exp \left(-\tfrac{ 1}{2}\Pe\varsigma u - \phi \varsigma |u| \right)$.
Therefore 
\begin{gather*}
\begin{split}
\int_{\mathcal{D}_1} \mathcal{G}_1(x_1'-x_1'') C_H(x_1'') F^{(1)}_{\sqrt{2}\varsigma}(x_1' - x_1'') \, \mathrm{d}x_1'' \approx 
- \dfrac{1}{4 \sqrt{\pi} \phi} C_H(x_1') \int_{-\infty}^{\infty} \exp \left(- \dfrac{u^2}{4} - \dfrac{\Pe}{2}\varsigma u - \phi \varsigma |u| \right)\, \mathrm{d}u.
\end{split}
\end{gather*}
The integral asymptotes to $2\sqrt{\pi}$ as $\varsigma\rightarrow 0$, 
and we obtain $\beta_1=1/(2\phi)$ in (\ref{eq:mc2}).
In 2D, (\ref{Free-Space-GF-2D_near_field}a) with $\mathbf{x}''=\mathbf{x}' + \varsigma \mathbf{u}$ and $\hat{r} = |\mathbf{u}|$ gives
$\mathcal{G}_2(\mathbf{x}'-\mathbf{x}'') = \mathcal{G}_2(-\varsigma \mathbf{u}) \approx ({1}/(2 \pi)) \log(\phi \varsigma \hat{r})$
when $\varsigma \ll 1/\phi \ll 1$. Therefore 
\begin{align*}
\int_{\mathcal{D}_2} \mathcal{G}_2(\mathbf{x}'-\mathbf{x}'') C_H(x_1'') F^{(2)}_{\sqrt{2}\varsigma}(\mathbf{x}' - \mathbf{x}'') \, \mathrm{d}\mathbf{x}'' 
&\approx \dfrac{1}{4\pi} C_H(x_1'') \int_0^{\infty} \hat{r} \log(\phi \varsigma \hat{r}) \exp \left(-\dfrac{\hat{r}^2}{4}\right) \, \mathrm{d} \hat{r}.
\end{align*}
The integral is evaluated using the identity
\begin{gather}
\label{eq:id}
\int_{0}^{\infty} x \log(bx) \exp \left(-ax^2\right) \mathrm{d}x = \dfrac{1}{2a} \log \left(\dfrac{b}{\sqrt{a}}\right) + \dfrac{\gamma}{4a}
\end{gather}
(using \citet{VanHeemert1957Cyclic}, where $\gamma \approx 0.577$ is the Euler--Mascheroni constant), to obtain 
$\beta_2=(\gamma-2\log(2\phi\varsigma))/(4\pi)$ in (\ref{eq:mc2}).
In 3D, (\ref{Free-Space-GF-3D_near_field}) with $\mathbf{x}''=\mathbf{x}' + \varsigma \mathbf{u}$ and $\hat{r} = |\mathbf{u}|$ gives
$\mathcal{G}_3(\mathbf{x}'-\mathbf{x}'') = \mathcal{G}_3(-\varsigma \mathbf{u}) \approx - {1}/(4 \pi \varsigma \hat{r})$
when $\varsigma \ll 1/\phi \ll 1$. Therefore 
\begin{align*}
\int_{\mathcal{D}_3} \mathcal{G}_3(\mathbf{x}'-\mathbf{x}'') C_H(x_1'') F^{(3)}_{\sqrt{2}\varsigma}(\mathbf{x}' - \mathbf{x}'') \, \mathrm{d}\mathbf{x}'' 
&\approx - \dfrac{1}{(4\pi)^{3/2} \varsigma} C_H(x_1'') \int_0^{\infty} \hat{r} \exp \left(-\dfrac{\hat{r}^2}{4}\right) \, \mathrm{d} \hat{r},
\end{align*}
giving $\beta_3=1/4 \pi^{3/2}\varsigma$ in (\ref{eq:mc2}).

Integrals involving the 1D Green's function convolved with $C_H$ can be evaluated exactly when the free-space function $\mathcal{G}_1$ is used.  These can be simplified by eliminating terms that are exponentially small throughout the domain, when $\Pe\gg 1$.  The resulting expressions are 
\begin{equation}\label{Int_GCH}
\int_{\mathcal{D}_1} \mathcal{G}_1(x_1 - x_1') C_H(x_1') \, \mathrm{d}x_1' \approx \dfrac{\Pe e^{\frac{\Pe}{2}x_1}}{4\phi^2 \hat{\psi}(1)} \Bigg( 2\Pe\, e^{\phi(x_1-1)}  - \left(2\phi+\Pe\right)\left(1+2\phi x_1\right) e^{\phi(1-x_1)} \Bigg),
\end{equation}
\begin{gather}
\begin{split}
\int_{\mathcal{D}_1}\int_{\mathcal{D}_1} \mathcal{G}_1(x_1 - x_1')\mathcal{G}_1(x_1' - x_1'') C_H(x_1'') \, \mathrm{d}x_1'\, \mathrm{d}x_1'' \approx& \frac{\Pe e^{\frac{\Pe}{2}x_1}}{8\phi^4 \hat{\psi}(1)} \Bigg( \left(2\phi+\Pe\right) (1+\phi x_1)^2e^{\phi(1-x_1)} \\
&+ \left( - \dfrac{5\Pe}{2} - \phi(1+\Pe+2\phi)\right)e^{\phi(x_1-1)} \Bigg)
\end{split}
\end{gather}
and
\begin{align}
\begin{split}
\int_{\mathcal{D}_1} [\mathcal{G}_1(x_1-x_1')C_H(x_1')]^2&  \, \mathrm{d}x_1' \approx  \left(\dfrac{\Pe^2}{16\phi^3\hat{\psi}(1)^2}\right) e^{\Pe x_1} \Bigg( \left(2\phi + \Pe\right)^2(4\phi x_1 + 1) e^{2\phi(1-x_1)} \\
&+ 4\left(4\phi^2 - \Pe^2\right)\left(2 - e^{-2\phi x_1}\right) - 4\left(4\phi^2 + 2\phi \Pe - \Pe^2 \right) e^{2\phi(x_1-1)} \Bigg),
\end{split}
\end{align}
with $\hat{\psi}$ being the approximation of $\psi$ near $x_1=1$, which is given by $\hat{\psi}(x_1) = (2\Pe\phi + \Pe^2 + 2\Da)e^{\phi x_1}$.

\section{Integrals for effective uptake}
\label{app:W}
\renewcommand{\theequation}{E.\arabic{equation}}

Consider a Gaussian covariance function of the form
$\widehat{\mathcal{K}}(\mathbf{x}-\mathbf{y}) = \sigma^2 \exp\left(-{|\mathbf{x}-\mathbf{y}|^2}/{\ell^2}\right)$.
Then (\ref{relation}) gives
\begin{equation*}
\mathcal{G}_2\widehat{\mathcal{K}}(\mathbf{0}) \approx - \dfrac{\sigma^2}{2 \pi} \int_{\mathbb{R}^2} \exp \left(\dfrac{\Pe}{2}x_1 - \dfrac{|\mathbf{x}|^2}{\ell^2} \right) K_0(\phi |\mathbf{x}|) \, \mathrm{d}\mathbf{x}.
\end{equation*}
By converting to polar coordinates where $x_1 = r \cos \theta$ and $x_2= r \sin \theta$,
we can solve the $\theta$ integral by using
\begin{equation}\label{thetaIntegral}
\int_0^{2\pi} \exp \left( z \cos \theta \right) \, \mathrm{d}\theta 
= 2 \pi I_0(z),
\end{equation}
where $I_{\nu}$ is a modified Bessel function of the first kind \citep{Abramowitz1964Handbook}, 
to give
\begin{equation*}
\mathcal{G}_2\widehat{\mathcal{K}}(\mathbf{0}) \approx - \sigma^2 \int_0^{\infty} \exp \left( - \dfrac{r^2}{\ell^2} \right) I_0 \left(\dfrac{\Pe}{2} r \right) K_0(\phi r) r \, \mathrm{d}r.
\end{equation*}
We set $r = \ell R$ and approximate the Bessel functions using 
\begin{equation}\label{besselExpansions}
I_0 \left((\Pe/2) \ell R \right) \approx 1 + \mathcal{O}(l^2 \Pe^2 R^2), \quad K_0(\phi \ell R) \approx - \log( \phi \ell R) = - \log(\phi \ell) - \log(R) \quad\mathrm{as}\quad \ell \rightarrow 0,
\end{equation}
to give
\begin{equation*}
\mathcal{G}_2\widehat{\mathcal{K}}(\mathbf{0}) \approx \sigma^2 \ell^2 \left( \log (\phi \ell) \int_0^{\infty} R \exp \left( - R^2 \right) \, \mathrm{d}R + \int_0^{\infty} R \log R \exp \left( - R^2 \right) \, \mathrm{d}R\right),
\end{equation*}
for suitably small $\ell$. Using (\ref{eq:id}) we obtain
$\mathcal{G}_2\widehat{\mathcal{K}}(\mathbf{0}) \approx - \tfrac{1}{4} {\sigma^2 \ell^2} \left( \gamma -  2\log (\phi \ell) \right)$,
hence yielding the 2D result in (\ref{Effective_Uptake_G_2D}).  The analogous integral in 3D reduces to 
\begin{equation*}
\mathcal{G}_3\widehat{\mathcal{K}}(\mathbf{0}) \approx - \dfrac{\sigma^2}{2} \int_{0}^{\pi} \int_{0}^{\infty} \exp \left( - \phi r - \dfrac{r^2}{\ell^2} \right) I_0 \left(\dfrac{\Pe}{2}r \sin \theta \right) r \sin \theta \, \mathrm{d}r\, \mathrm{d}\theta.
\end{equation*}
Again for small $\ell$ we use (\ref{besselExpansions}) to evaluate the integral for $r=O(\ell)$, 
leading to 
$\mathcal{G}_3\widehat{\mathcal{K}}(\mathbf{0}) \approx - \tfrac{1}{2}{\sigma^2 \ell^2}$, 
the 3D limit in (\ref{Effective_Uptake_G_2D}).

\section*{Acknowledgements}
OEJ and ILC acknowledge support from EPSRC grant EP/T008725/1.

{For the purpose of open access, the authors have applied a Creative Commons Attribution (CC BY) licence to any Author Accepted Manuscript version arising.}


\bibliography{References.bib}
\bibliographystyle{apalike} 

\end{document}